\documentclass[conference]{IEEEtran}
\IEEEoverridecommandlockouts
\usepackage{lineno}
\usepackage{cite}
\usepackage{amsmath,amssymb,amsfonts}
\usepackage{amsmath}
\usepackage{amsthm}
\DeclareMathOperator*{\argmax}{argmax}

\usepackage{graphicx}
\usepackage{textcomp}
\usepackage{xcolor}
\usepackage{graphicx}
\usepackage{float}
\usepackage{subfigure}
\usepackage{amsmath}
\usepackage{amsfonts,amssymb}
\usepackage{mathrsfs}
\usepackage{mathtools}
\usepackage{algpseudocode}
\usepackage{bm}
\usepackage{multirow}
\usepackage{array}
\usepackage{amssymb}
\usepackage{amsmath}
\usepackage{cite}
\usepackage{url}
\usepackage{xcolor}
\usepackage{cite,graphicx,amsmath,amssymb}
\usepackage{subfigure}
\usepackage{fancyhdr}
\usepackage{mdwmath}
\usepackage{mdwtab}
\usepackage{caption}
\usepackage{amsthm}
\usepackage{setspace}
\usepackage{bm}
\usepackage{algorithm}
\usepackage{algpseudocode}
\usepackage{mathtools}
\usepackage{dsfont}
\usepackage{bbm}
\usepackage{multicol}
\usepackage{framed}
\newtheorem{remark}{Remark}
\newtheorem{theorem}{Theorem}

\newtheorem{lemma}{Lemma}

\newtheorem{corollary}{Corollary}

\makeatletter
\newcommand{\biggg}{\bBigg@{3}}
\newcommand{\Biggg}{\bBigg@{3.5}}
\makeatother
\def\BibTeX{{\rm B\kern-.05em{\sc i\kern-.025em b}\kern-.08em
    T\kern-.1667em\lower.7ex\hbox{E}\kern-.125emX}}
\begin{document}
\title{On the Performance of Continuous Aperture Array (CAPA)-Based Wireless Communications}
\author{\IEEEauthorblockN{Chongjun~Ouyang$^{\star\ddag}$, Yuanwei~Liu$^{\star}$, and Xingqi~Zhang$^{\dag}$}
$^{\star}$School of Electronic Engineering and Computer Science, Queen Mary University of London\\
$^\ddag$School of Electrical and Electronic Engineering, University College Dublin\\
$^\dag$Department of Electrical and Computer Engineering, University of Alberta\\
Email: $^{\star}$\{c.ouyang, yuanwei.liu\}@qmul.ac.uk, $^{\dag}$xingqi.zhang@ualberta.ca}
\maketitle
\begin{abstract}
The performance of continuous aperture array (CAPA)-based wireless communications is analyzed in an uplink scenario. An analytical framework is proposed to characterize uplink CAPA-based transmission using electromagnetic field theories. On this basis, new expressions are derived for the channel capacity in a single-user scenario and the sum-rate capacity in a multiuser scenario, along with the capacity-achieving decoding schemes. These findings are proved to differ greatly from those established for conventional spatially discrete (SPD) arrays. Numerical results are provided to demonstrate that CAPA offers significant capacity gains compared to the SPD array.
\end{abstract} 
\begin{IEEEkeywords}
Channel capacity, continuous aperture array (CAPA), performance analysis.
\end{IEEEkeywords}
\section{Introduction}
Multiple antenna technology serves as a cornerstone in the advancement of modern wireless communication systems. At its core lies the principle of leveraging an increased number of antenna elements to enhance spatial degrees of freedom (DoFs) and augment channel capacity \cite{tse2005fundamentals}.

Traditionally, multiple antenna prototypes adhere to a spatially discrete (SPD) topological structure, wherein each antenna is represented as an individual point \cite{liu2023near}. Within a fixed array aperture, decreasing the inter-element distance facilitates the deployment of additional antennas, thereby enhancing spatial DoFs and giving rise to the concept of holographic arrays. In holographic arrays, SPD antenna elements are arranged in a sub-half-wavelength-spaced manner \cite{pizzo2020spatially}.

This paper introduces the concept of a \emph{continuous aperture array (CAPA)}, where the entire array forms a spatially-continuous electromagnetic (EM) aperture. In essence, a CAPA can be conceptualized as an SPD array comprising an \emph{infinite number} of antennas with \emph{infinitesimal spacing} \cite{liu2023near,liaskos2018new}. CAPA holds the potential to offer more spatial DoFs compared to SPD arrays, thereby further enhancing system performance. 

In contrast to traditional arrays that rely on spatially discrete modeling, the modeling of CAPA-based wireless transmission is grounded in \emph{EM fields}. Specifically, while channels for a traditional SPD array are typically modeled using discrete matrices, the spatial response for a CAPA should be represented by a \emph{continuous operator in the Hilbert space} \cite{liu2023near}. This fundamental difference renders the traditional SPD arrays-based transmission framework inapplicable to CAPAs. Currently, research on CAPAs is still in its infancy, with most existing works limited to single-user scenarios; see \cite{liu2023near,liu2024near} for more details. Although some authors have explored multiuser cases \cite{zhang2023pattern,xu2023exploiting}, these efforts have not fully characterized the fundamental performance limits of CAPAs. Furthermore, an analytically tractable framework for CAPA-based communications is still missing.

Motivated by these knowledge gaps, we analyze the performance of CAPA-based communications by investigating an uplink system. The main contributions of this article are summarized as follows: {\romannumeral1}) We propose an uplink transmission framework for CAPAs leveraging EM field theories. {\romannumeral2}) We derive the channel capacity of a single-user uplink CAPA system and design the optimal detector required to achieve it. {\romannumeral3}) We extend our analysis to a multiuser uplink CAPA system, derive the sum-rate capacity, and identify the decoding rule necessary for achieving it. {\romannumeral4}) We specialize the derived results to several special aperture structures, including the SPD array, and prove the capacity achieved by an SPD array is limited by its array occupation ratio. {\romannumeral5}) We provide numerical results to demonstrate that CAPAs offer significant capacity enhancements over traditional SPD arrays and that as the array occupation ratio of an SPD array approaches $1$, its capacity converges to that of a CAPA.
\section{System Model}\label{Section: System Model}
\begin{figure}[!t]
 \centering
\setlength{\abovecaptionskip}{0pt}
\includegraphics[height=0.3\textwidth]{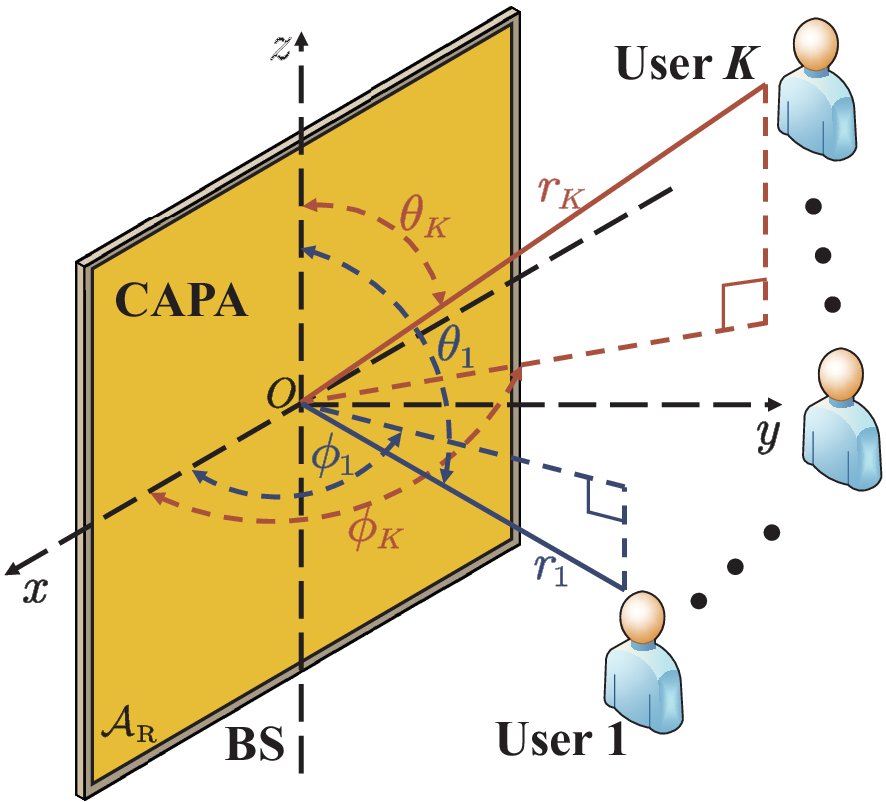}
\caption{Illustration of a CAPA-based multiuser channel.}
\label{Figure: System_Model}
\end{figure}

We study an uplink channel where $K$ users simultaneously transmit signals to a base station (BS) equipped with a CAPA, as shown in {\figurename} {\ref{Figure: System_Model}}. The CAPA is centered at the origin $O\triangleq[0,0,0]^{\mathsf{T}}$, and its aperture is denoted as $\mathcal{A}_{\mathsf{R}}\subseteq{\mathbbmss{R}}^{3\times1}$. Let $r_k$ denote the distance from the center of the CAPA to the center location of each user $k\in{\mathcal{K}}\triangleq\{1,\ldots,K\}$, and $\phi_k\in[0,\pi]$ and $\theta_k\in[0,\pi]$ denote the associated azimuth and elevation angles, respectively. Therefore, user $k$'s center location can be expressed as ${\mathbf{s}}_{k}=[r_k\Phi_k,r_k\Psi_k,r_k\Theta_k]^{\mathsf{T}}$, where $\Phi_k\triangleq\cos{\phi_k}\sin{\theta_k}$, $\Psi_k\triangleq\sin{\phi_k}\sin{\theta_k}$, and $\Theta_k\triangleq\cos{\theta_k}$. Assume that user $k$ employs a hypothetical isotropic antenna for signal transmission, whose transmit aperture is denoted as ${\mathcal{A}}_{k}$ with $\lvert{\mathcal{A}}_{k}\rvert=\frac{\lambda^2}{4\pi}$ for $k\in{\mathcal{K}}$, where $\lambda$ denotes the wavelength. For clarity, we assume that the aperture size of ${\mathcal{A}}_{\mathsf{R}}$ significantly exceeds that of ${\mathcal{A}}_{k}$ ($\forall k$), i.e., $\lvert{\mathcal{A}}_{\mathsf{R}}\rvert\gg \lvert{\mathcal{A}}_{k}\rvert$. 
\subsection{Signal Model}
Let us define ${\mathsf{J}}_k({\mathbf{s}})\in{\mathbbmss{C}}$ as the continuous distribution of source currents generated by user $k$ to convey his data information, where ${\mathbf{s}}\in{\mathcal{A}}_{k}$. The excited electric field ${\mathsf{E}}_k(\mathbf{r})\in{\mathbbmss{C}}$ due to ${\mathsf{J}}_k({\mathbf{s}})$ at point $\mathbf{r}\in{\mathcal{A}}_{\mathsf{R}}$ can be expressed as follows \cite{balanis2016antenna}:
{\setlength\abovedisplayskip{2pt}
\setlength\belowdisplayskip{2pt}
\begin{align}\label{User_k_electric_radiation_field}
{\mathsf{E}}_k(\mathbf{r})=\int_{{\mathcal{A}}_{k}}{\mathsf{h}}(\mathbf{r},{\mathbf{s}})s_k{\mathsf{J}}_k({\mathbf{s}}){\rm{d}}{\mathbf{s}},
\end{align}
}where $s_k\in{\mathbbmss{C}}$ represents the normalized data symbol subject to ${\mathbbmss{E}}\{\lvert s_k\rvert^2\}=1$, and ${\mathsf{h}}(\mathbf{r},{\mathbf{s}})\in{\mathbbmss{C}}$ denotes the spatial channel response between $\mathbf{s}$ and $\mathbf{r}$. Since $\lvert{\mathcal{A}}_{\mathsf{R}}\rvert\gg \lvert{\mathcal{A}}_{k}\rvert$, we simplify ${\mathsf{E}}_k(\mathbf{r})$ as follows:
{\setlength\abovedisplayskip{2pt}
\setlength\belowdisplayskip{2pt}
\begin{align}\label{Simplified_electric_radiation_field}
{\mathsf{E}}_k(\mathbf{r})\approx s_k{\mathsf{h}}(\mathbf{r},{\mathbf{s}}_{k}){\mathsf{J}}_k({\mathbf{s}}_{k})\lvert{\mathcal{A}}_{k}\rvert.
\end{align}
}The total observed electric field ${\mathsf{Y}}(\mathbf{r})\in{\mathbbmss{C}}$ at point $\mathbf{r}\in{\mathcal{A}}_{\mathsf{R}}$ is the sum of the information-carrying electric fields $\{{\mathsf{E}}_k(\mathbf{r})\}_{k=1}^{K}$, along with a random noise field ${\mathsf{N}}(\mathbf{r})\in{\mathbbmss{C}}$, i.e.,
{\setlength\abovedisplayskip{2pt}
\setlength\belowdisplayskip{2pt}
\begin{subequations}\label{Total_electric_radiation_field}
\begin{align}
{\mathsf{Y}}(\mathbf{r})&=\sum\nolimits_{k=1}^{K}{\mathsf{E}}_k(\mathbf{r})+{\mathsf{N}}(\mathbf{r})\\
&\approx\sum\nolimits_{k=1}^{K}s_k{\mathsf{h}}(\mathbf{r},{\mathbf{s}}_{k}){\mathsf{J}}_k({\mathbf{s}}_{k})\lvert{\mathcal{A}}_{k}\rvert
+{\mathsf{N}}(\mathbf{r}),
\end{align}
\end{subequations}
}where ${\mathsf{N}}(\mathbf{r})$ accounts for the thermal noise. The noise field is modeled as a zero-mean complex Gaussian random process satisfying ${\mathbbmss{E}}\{{\mathsf{N}}(\mathbf{r}_1){\mathsf{N}}^{\mathsf{H}}(\mathbf{r}_2)\}={\sigma}^2\delta(\mathbf{r}_1-\mathbf{r}_2)$, where $\delta(\cdot)$ represents the Dirac delta function. ${\mathsf{N}}(\mathbf{r})$ and $\{s_k\}_{k=1}^{K}$ are assumed to be uncorrelated.

After observing ${\mathsf{Y}}(\mathbf{r})$, the BS should employ a properly designed detector along with maximum-likelihood (ML) decoding to recover the data information contained in $\{s_k\}_{k=1}^{K}$, which shall be detailed in the following pages.
\subsection{Channel Model}
To facilitate theoretical investigations into fundamental performance limits and asymptotic behaviors, we focus our discussion on line-of-sight (LoS) channels. Under this consideration, ${\mathsf{h}}(\mathbf{r},{\mathbf{s}})$ is modeled as follows \cite{ouyang2024reactive}:
{\setlength\abovedisplayskip{2pt}
\setlength\belowdisplayskip{2pt}
\begin{align}\label{CAPA_LoS_Channel_Model}
{\mathsf{h}}(\mathbf{r},{\mathbf{s}})={\mathsf{h}}_{\mathsf{em}}(\mathbf{r},{\mathbf{s}})
{\mathsf{h}}_{\mathsf{pa}}(\mathbf{r},{\mathbf{s}}).
\end{align}
}The terms appearing in \eqref{CAPA_LoS_Channel_Model} are defined as follows:
\begin{itemize}
  \item ${\mathsf{h}}_{\mathsf{pa}}(\mathbf{r},{\mathbf{s}})\triangleq\sqrt{\frac{\lvert{\mathbf{e}}_{{\mathbf{r}}}^{\mathsf{T}}({\mathbf{s}}-{\mathbf{r}})\rvert}{\lVert{\mathbf{r}}-{\mathbf{s}}\rVert}}$ models the impact of the projected aperture of the BS array, which is reflected by the projection of the normal vector onto the wave propagation direction at each local point $\mathbf{r}$.
  \item ${\mathbf{e}}_{{\mathbf{r}}}\in{\mathbbmss{R}}^{3\times1}$ is the normal vector of the CAPA at $\mathbf{r}\in{\mathcal{A}}_{\mathsf{R}}$.
  \item The function
  {\setlength\abovedisplayskip{2pt}
\setlength\belowdisplayskip{2pt}
\begin{align}\label{Green_Function_Full_Version}
{\mathsf{h}}_{\mathsf{em}}(\mathbf{r},{\mathbf{s}})\triangleq\frac{{\rm{j}}k_0\eta{\rm{e}}^{-{\rm{j}}k_0\lVert{\mathbf{r}}-{\mathbf{s}}\rVert}
}{4\pi \lVert{\mathbf{r}}-{\mathbf{s}}\rVert}\left(1+\frac{{\rm{j}}/k_0}{\lVert{\mathbf{r}}-{\mathbf{s}}\rVert}-
\frac{1/k_0^2}{\lVert{\mathbf{r}}-{\mathbf{s}}\rVert^2}\right)
\end{align}
}models the influence of free-space EM propagation, where $\eta=120\pi$ is the impedance of free space and $k_0=\frac{2\pi}{\lambda}$ is the wavenumber.
\end{itemize}
The function ${\mathsf{h}}_{\mathsf{em}}(\mathbf{r},{\mathbf{s}})$ comprises three terms: the first term corresponds to the radiating near-field and far-field regions, while the remaining two terms correspond to the reactive near-field region. The squared magnitude of the parenthesis in \eqref{Green_Function_Full_Version} is $1-\frac{1/k_0^2}{\lVert{\mathbf{r}}-{\mathbf{s}}\rVert^2}+\frac{1/k_0^4}{\lVert{\mathbf{r}}-{\mathbf{s}}\rVert^4}$, which equals $0.97$ at distance $\lVert{\mathbf{r}}-{\mathbf{s}}\rVert=\lambda$. Hence, when considering practical systems with $r_k\gg \lambda$, the last two terms in \eqref{Green_Function_Full_Version} can be neglected; see \cite{ouyang2024reactive} for a more detailed theoretical justification. We thus neglect the reactive terms and approximate \eqref{Green_Function_Full_Version} as ${\mathsf{h}}_{\mathsf{em}}(\mathbf{r},{\mathbf{s}})\approx\frac{{\rm{j}}k_0\eta{\rm{e}}^{-{\rm{j}}k_0\lVert{\mathbf{r}}-{\mathbf{s}}\rVert}
}{4\pi \lVert{\mathbf{r}}-{\mathbf{s}}\rVert}$.
\section{Single-User Case}
Having established the system model, we next analyze the performance of CAPA-based uplink communications. For clarity, we commence with the single-user case.

By setting $K=1$ and dropping the user index $k$, the observed electric field degenerates into
{\setlength\abovedisplayskip{2pt}
\setlength\belowdisplayskip{2pt}
\begin{align}\label{Single_User_electric_radiation_field}
{\mathsf{Y}}(\mathbf{r})=s{\mathsf{h}}(\mathbf{r},{\mathbf{s}}){\mathsf{J}}_k({\mathbf{s}})\lvert{\mathcal{A}}\rvert
+{\mathsf{N}}(\mathbf{r}).
\end{align}
}We then design a detector ${\mathsf{v}}(\mathbf{r})$ to recover the data information contained in $s$. Applying this detector to ${\mathsf{Y}}(\mathbf{r})$ yields
{\setlength\abovedisplayskip{2pt}
\setlength\belowdisplayskip{2pt}
\begin{equation}\label{Single_User_Detection}
\begin{split}
\int_{{\mathcal{A}}_{\mathsf{R}}}{\mathsf{v}}^{\mathsf{H}}(\mathbf{r}){\mathsf{Y}}(\mathbf{r}){\rm{d}}{\mathbf{r}}
&=s{\mathsf{J}}_k({\mathbf{s}})\lvert{\mathcal{A}}\rvert\int_{{\mathcal{A}}_{\mathsf{R}}}{\mathsf{v}}^{\mathsf{H}}(\mathbf{r}){\mathsf{h}}(\mathbf{r},{\mathbf{s}}){\rm{d}}{\mathbf{r}}\\
&+\int_{{\mathcal{A}}_{\mathsf{R}}}{\mathsf{v}}^{\mathsf{H}}(\mathbf{r}){\mathsf{N}}(\mathbf{r}){\rm{d}}{\mathbf{r}}.
\end{split}
\end{equation}
}Since ${\mathsf{N}}(\mathbf{r})$ is a Gaussian random field, $\int_{{\mathcal{S}}_{\mathsf{R}}}{\mathsf{h}}^{\mathsf{H}}(\mathbf{r},{\mathbf{s}}_{\mathsf{u}}){\mathsf{N}}(\mathbf{r}){\rm{d}}{\mathbf{r}}$ is complex Gaussian distributed, whose mean and variance is calculated as follows. 
\vspace{-5pt}
\begin{lemma}\label{Lemma_Noise_Distribution}
Under the assumption that ${\mathsf{N}}(\mathbf{r})$ is a zero-mean complex Gaussian process with ${\mathbbmss{E}}\{{\mathsf{N}}(\mathbf{r}_1){\mathsf{N}}^{\mathsf{H}}(\mathbf{r}_2)\}={\sigma}^2\delta(\mathbf{r}_1-\mathbf{r}_2)$, we obtain $\int_{{\mathcal{S}}_{\mathsf{R}}}{\mathsf{h}}^{\mathsf{H}}(\mathbf{r},{\mathbf{s}}_{\mathsf{u}}){\mathsf{N}}(\mathbf{r}){\rm{d}}{\mathbf{r}}\sim{\mathcal{CN}}(0,\sigma^2{\mathsf{a}}_{\mathsf{R}})$.
\end{lemma}
\vspace{-5pt}
\begin{IEEEproof}
Please refer to Appendix \ref{Proof_Lemma_Noise_Distribution} for more details.
\end{IEEEproof}
Taken together, the single-to-noise ratio (SNR) for decoding $s$ is given by
{\setlength\abovedisplayskip{2pt}
\setlength\belowdisplayskip{2pt}
\begin{equation}\label{Single_User_SNR}
\begin{split}
\gamma=
\frac{\lvert{\mathsf{J}}({\mathbf{s}})\rvert^2\lvert{\mathcal{A}}\rvert^2
\lvert\int_{{\mathcal{A}}_{\mathsf{R}}}{\mathsf{v}}^{\mathsf{H}}(\mathbf{r}){\mathsf{h}}(\mathbf{r},{\mathbf{s}}){\rm{d}}{\mathbf{r}}\rvert^2}
{\sigma^2\int_{{\mathcal{A}}_{\mathsf{R}}}\lvert{\mathsf{v}}(\mathbf{r})\rvert^2{\rm{d}}{\mathbf{r}}}.
\end{split}
\end{equation}
}The subsequent task is to find a detector that can maximize the SNR shown in \eqref{Single_User_SNR}. By observing the mathematical structure of $\gamma$, we note that $\gamma$ is independent of the norm of ${\mathsf{v}}(\mathbf{r})$, i.e., $\int_{{\mathcal{A}}_{\mathsf{R}}}\lvert{\mathsf{v}}(\mathbf{r})\rvert^2{\rm{d}}{\mathbf{r}}$. On this basis, the problem of designing ${\mathsf{v}}(\mathbf{r})$ is equivalently transformed as follows:
{\setlength\abovedisplayskip{2pt}
\setlength\belowdisplayskip{2pt}
\begin{equation}\label{Optimal_Detector_MISO_SU_Problem}
\argmax\nolimits_{{\mathsf{v}}(\mathbf{r})}{\left\lvert\int_{{\mathcal{A}}_{\mathsf{R}}}{\mathsf{v}}^{\mathsf{H}}(\mathbf{r}){\mathsf{h}}(\mathbf{r},{\mathbf{s}}){\rm{d}}{\mathbf{r}}
\right\rvert^2},
\end{equation}
}whose optimal solution, i.e., the optimal detector, is given by
{\setlength\abovedisplayskip{2pt}
\setlength\belowdisplayskip{2pt}
\begin{equation}\label{Optimal_Detector_MISO_SU_Answer}
{\mathsf{v}}(\mathbf{r})={\mathsf{h}}(\mathbf{r},{\mathbf{s}}).
\end{equation}
}\vspace{-5pt}
\begin{remark}
The result in \eqref{Optimal_Detector_MISO_SU_Answer} suggests that the capacity-achieving detector for single-user CAPA communications aligns with the spatial channel response ${\mathsf{h}}(\mathbf{r},{\mathbf{s}})$, which can be regarded as continuous version of the maximal-ratio combiner.  
\end{remark}
\vspace{-5pt}
Substituting \eqref{Optimal_Detector_MISO_SU_Answer} back into \eqref{Single_User_SNR} gives
{\setlength\abovedisplayskip{2pt}
\setlength\belowdisplayskip{2pt}
\begin{equation}\label{Single_User_SNR}
\begin{split}
\gamma=\frac{\lvert{\mathsf{J}}({\mathbf{s}})\rvert^2\lvert{\mathcal{A}}\rvert^2}{\sigma^2}
\int_{{\mathcal{A}}_{\mathsf{R}}}\lvert{\mathsf{h}}(\mathbf{r},{\mathbf{s}})\rvert^2{\rm{d}}{\mathbf{r}}
=\overline{\gamma}
\int_{{\mathcal{A}}_{\mathsf{R}}}\lvert{\mathsf{g}}(\mathbf{r},{\mathbf{s}})\rvert^2{\rm{d}}{\mathbf{r}},
\end{split}
\end{equation}
}where $\overline{\gamma}\triangleq\frac{\lvert{\mathsf{J}}({\mathbf{s}})\rvert^2\lvert{\mathcal{A}}\rvert^2k_0^2\eta^2}{4\pi\sigma^2}$ and ${\mathsf{g}}(\mathbf{r},{\mathbf{s}})\triangleq\frac{{\rm{e}}^{-{\rm{j}}k_0\lVert{\mathbf{r}}-{\mathbf{s}}\rVert}}{\sqrt{4\pi} \lVert{\mathbf{r}}-{\mathbf{s}}\rVert}
{\mathsf{h}}_{\mathsf{pa}}(\mathbf{r},{\mathbf{s}})$. Note that $\frac{{\rm{e}}^{-{\rm{j}}k_0\lVert{\mathbf{r}}-{\mathbf{s}}\rVert}}{\sqrt{4\pi} \lVert{\mathbf{r}}-{\mathbf{s}}\rVert}$ characterizes the spherical-wave propagation in arbitrary homogeneous mediums, and thus we term $\int_{{\mathcal{A}}_{\mathsf{R}}}\lvert{\mathsf{g}}(\mathbf{r},{\mathbf{s}})\rvert^2{\rm{d}}{\mathbf{r}}\triangleq {\mathsf{a}}_{\mathsf{R}}$ as the array gain. The remaining term $\overline{\gamma}$ is termed as the transmit SNR.

Building upon \eqref{Single_User_SNR}, the channel capacity under the single-user scenario can be written as follows:
{\setlength\abovedisplayskip{2pt}
\setlength\belowdisplayskip{2pt}
\begin{align}\label{SU_Channel_Capacity}
{\mathsf{C}}=\log_2(1+\gamma)=\log_2(1+{\overline{\gamma}}{\mathsf{a}}_{\mathsf{R}}).
\end{align}
}We note that the channel capacity is mainly influenced by the receive aperture $\mathsf{A}_{\mathsf{R}}$ as well as the user's center location $\mathbf{s}$. 
\section{Multiuser Case}\label{Section: CAP Arrays}
This section extends the single-user scenario to the more general multiuser scenario, with the \emph{sum-rate capacity} as the performance metric of interest. For brevity, our analyses focus on the two-user case. The extension to scenarios with an arbitrary number of users will be left to future work.
\subsection{Optimal Multiuser Detection}
When $K=2$, the observed electric field in \eqref{Total_electric_radiation_field} can be rewritten as follows:
{\setlength\abovedisplayskip{2pt}
\setlength\belowdisplayskip{2pt}
\begin{equation}\label{Two_User_Total_electric_radiation_field}
\begin{split}
{\mathsf{Y}}(\mathbf{r})
&=s_2{\mathsf{h}}(\mathbf{r},{\mathbf{s}}_{2}){\mathsf{J}}_2({\mathbf{s}}_{2})\lvert{\mathcal{A}}_{2}\rvert\\
&+s_1{\mathsf{h}}(\mathbf{r},{\mathbf{s}}_{1}){\mathsf{J}}_1({\mathbf{s}}_{1})\lvert{\mathcal{A}}_{1}\rvert+{\mathsf{N}}(\mathbf{r}).
\end{split}
\end{equation}
}The sum-rate capacity of an uplink multiuser channel can be achieved by using \emph{successive interference cancellation (SIC) decoding} \cite{tse2005fundamentals}. More specifically, the message sent by one user is first decoded by treating the message from the other user as interference and then removed, and the other message is decoded without inter-user interference (IUI).

Given the uplink signal model \eqref{Two_User_Total_electric_radiation_field}, there exist two SIC orders: $1\rightarrow2$ and $2\rightarrow1$. In the sequel, we discuss the sum-rate capacity achieved by the decoding order $2\rightarrow1$. In this case, $s_2$ is decoded from ${\mathsf{Y}}(\mathbf{r})$ by treating $s_1{\mathsf{h}}(\mathbf{r},{\mathbf{s}}_{1}){\mathsf{J}}_1({\mathbf{s}}_{1})\lvert{\mathcal{A}}_{1}\rvert$ as interference. After $s_2$ is decoded and subtracted from ${\mathsf{Y}}(\mathbf{r})$, $s_1$ is decoded without IUI from the following signal:
{\setlength\abovedisplayskip{2pt}
\setlength\belowdisplayskip{2pt}
\begin{align}\label{Two_User_SIC_electric_radiation_field}
{\mathsf{Z}}(\mathbf{r})\triangleq s_1{\mathsf{h}}(\mathbf{r},{\mathbf{s}}_{1}){\mathsf{J}}_1({\mathbf{s}}_{1})\lvert{\mathcal{A}}_{1}\rvert+{\mathsf{N}}(\mathbf{r}),
\end{align}
}which is the same as that given in \eqref{Single_User_electric_radiation_field}. Following the derivation steps in obtaining \eqref{Optimal_Detector_MISO_SU_Answer}, we know that the BS can use ${\mathsf{h}}(\mathbf{r},{\mathbf{s}}_{1})$ as the detector to recover $s_1$ from \eqref{Two_User_SIC_electric_radiation_field}. As a result, the SNR in decoding the message sent by user $1$ is given by
{\setlength\abovedisplayskip{2pt}
\setlength\belowdisplayskip{2pt}
\begin{align}
\gamma_1={\overline{\gamma}}_1{\mathsf{a}}_{1},
\end{align}
}where $\overline{\gamma}_1=\frac{\lvert{\mathsf{J}}_1({\mathbf{s}}_{1})\rvert^2\lvert{\mathcal{A}}_1\rvert^2k_0^2\eta^2}{4\pi\sigma^2}$ is the transmit SNR of user $1$ and ${\mathsf{a}}_{1}=\int_{{\mathcal{A}}_{\mathsf{R}}}\lvert{\mathsf{g}}(\mathbf{r},{\mathbf{s}}_1)\rvert^2{\rm{d}}{\mathbf{r}}$ is the associated channel gain. 

We then investigate the decoding of $s_2$ from ${\mathsf{Y}}(\mathbf{r})$, where the interference-plus-noise term is given by ${\mathsf{Z}}(\mathbf{r})$ shown in \eqref{Two_User_SIC_electric_radiation_field}. Since ${\mathsf{N}}(\mathbf{r})$ and $s_1$ are uncorrelated, the autocorrelation function of the random field ${\mathsf{Z}}(\mathbf{r})$ is given by
{\setlength\abovedisplayskip{2pt}
\setlength\belowdisplayskip{2pt}
\begin{equation}\label{Autocorrelatioon_Interference_Noise}
\begin{split}
{\mathbbmss{E}}\{{\mathsf{Z}}(\mathbf{r}_1){\mathsf{Z}}^{\mathsf{H}}(\mathbf{r}_2)\}&=
{\mathsf{g}}(\mathbf{r}_1,{\mathbf{s}}_{1}){\mathsf{g}}^{\mathsf{H}}(\mathbf{r}_2,{\mathbf{s}}_{1})
\overline{\gamma}_1\sigma^2\\
&+{\sigma}^2\delta(\mathbf{r}_1-\mathbf{r}_2)\triangleq {\mathscr{R}}_{{\mathsf{Z}}{\mathsf{Z}}}(\mathbf{r}_1,\mathbf{r}_2).
\end{split}
\end{equation}
}In order to obtain the optimal detector that maximizes the achievable rate of user $2$, we need to first design an invertible linear transformation ${\mathscr{K}}_{{\mathsf{Z}}}(\mathbf{r}',\mathbf{r})$ to whiten ${\mathsf{Z}}(\mathbf{r})$, i.e., finding a ${\mathscr{K}}_{{\mathsf{Z}}}(\mathbf{r}',\mathbf{r})$ that makes the autocorrelation function of $\int_{{\mathcal{A}}_{\mathsf{R}}}{\mathscr{K}}_{{\mathsf{Z}}}(\mathbf{r}',\mathbf{r}){\mathsf{Z}}(\mathbf{r}){\rm{d}}\mathbf{r}\triangleq {\mathsf{Z}}_{\mathsf{w}}(\mathbf{r}')$ proportional to the Dirac delta function. To this end, we introduce the following lemmas.
\vspace{-5pt}
\begin{lemma}\label{Lemma_IN_Invertible}
Let ${\mathscr{K}}_{{\mathsf{Z}}}(\mathbf{r}',\mathbf{r})=\delta(\mathbf{r}'-\mathbf{r})+\lambda
{\mathsf{g}}(\mathbf{r}',{\mathbf{s}}_{1}){\mathsf{g}}^{\mathsf{H}}(\mathbf{r},{\mathbf{s}}_{1})$ and $\overline{\mathscr{K}}_{{\mathsf{Z}}}(\mathbf{r}',\mathbf{r})=\delta(\mathbf{r}'-\mathbf{r})-\overline{\lambda}
{\mathsf{g}}(\mathbf{r}',{\mathbf{s}}_{1}){\mathsf{g}}^{\mathsf{H}}(\mathbf{r},{\mathbf{s}}_{1})$ with $\overline{\lambda}=\frac{\lambda}{1+\lambda {\mathsf{a}}_{1}}$. Then, it has
{\setlength\abovedisplayskip{2pt}
\setlength\belowdisplayskip{2pt}
\begin{equation}
\int_{{\mathcal{A}}_{\mathsf{R}}}\overline{\mathscr{K}}_{{\mathsf{Z}}}(\mathbf{r}'',\mathbf{r}')
\int_{{\mathcal{A}}_{\mathsf{R}}}{\mathscr{K}}_{{\mathsf{Z}}}(\mathbf{r}',\mathbf{r}){\mathsf{u}}(\mathbf{r})
{\rm{d}}\mathbf{r}{\rm{d}}\mathbf{r}'= {\mathsf{u}}(\mathbf{r}'')
\end{equation}
}for an arbitrary function ${\mathsf{u}}(\mathbf{r})$ defined in $\mathbf{r}\in{\mathcal{A}}_{\mathsf{R}}$.
\end{lemma}
\vspace{-5pt}
\begin{IEEEproof}
Please refer to Appendix \ref{Proof_Lemma_IN_Invertible} for more details.
\end{IEEEproof}
\vspace{-5pt}
\begin{lemma}\label{Lemma_IN_Whiten}
Let ${\mathscr{K}}_{{\mathsf{Z}}}(\mathbf{r}',\mathbf{r})=\delta(\mathbf{r}'-\mathbf{r})+\lambda
{\mathsf{g}}(\mathbf{r}',{\mathbf{s}}_{1}){\mathsf{g}}^{\mathsf{H}}(\mathbf{r},{\mathbf{s}}_{1})$ with $\lambda=-\frac{1}{{\mathsf{a}}_{1}}\pm\frac{1}{{\mathsf{a}}_{1}
\sqrt{1+\overline{\gamma}_{1}{\mathsf{a}}_{1}}}\triangleq \lambda^{\star}$. Then, the autocorrelation function of ${\mathsf{Z}}_{\mathsf{w}}(\mathbf{r}')$ satisfies ${\mathbbmss{E}}\{{\mathsf{Z}}_{\mathsf{w}}(\mathbf{r}_1){\mathsf{Z}}_{\mathsf{w}}^{\mathsf{H}}(\mathbf{r}_2)\}=
{\sigma}^2\delta(\mathbf{r}_1-\mathbf{r}_2)$.
\end{lemma}
\vspace{-5pt}
\begin{IEEEproof}
Please refer to Appendix \ref{Proof_Lemma_IN_Whiten} for more details.
\end{IEEEproof}
\vspace{-5pt}
\begin{remark}\label{Remark_Invertible_Linear_Transform}
The results in Lemma \ref{Lemma_IN_Invertible} suggest that ${\mathscr{K}}_{{\mathsf{Z}}}(\mathbf{r}',\mathbf{r})=\delta(\mathbf{r}'-\mathbf{r})+\lambda
{\mathsf{g}}(\mathbf{r}',{\mathbf{s}}_{1}){\mathsf{g}}^{\mathsf{H}}(\mathbf{r},{\mathbf{s}}_{1})$ is an invertible linear transformation. Applying this transformation to ${\mathsf{Y}}(\mathbf{r})$ is information lossless and has no influence on the channel capacity \cite{tse2005fundamentals}.
\end{remark}
\vspace{-5pt}
\vspace{-5pt}
\begin{remark}\label{Remark_IN_Whiten}
The results in Lemma \ref{Lemma_IN_Whiten} suggest that setting $\lambda=\lambda^{\star}$ makes ${\mathscr{K}}_{{\mathsf{Z}}}(\mathbf{r}',\mathbf{r})=\delta(\mathbf{r}'-\mathbf{r})+\lambda
{\mathsf{g}}(\mathbf{r}',{\mathbf{s}}_{1}){\mathsf{g}}^{\mathsf{H}}(\mathbf{r},{\mathbf{s}}_{1})$ capable of whitening the interference-plus-noise term $s_1{\mathsf{h}}(\mathbf{r},{\mathbf{s}}_{1}){\mathsf{J}}_1({\mathbf{s}}_{1})\lvert{\mathcal{A}}_{1}\rvert+{\mathsf{N}}(\mathbf{r})$.
\end{remark}
\vspace{-5pt}
Leveraging the insights unveiled in Remarks \ref{Remark_Invertible_Linear_Transform} and \ref{Remark_IN_Whiten}, we exploit ${\mathscr{K}}_{{\mathsf{Z}}}(\mathbf{r}',\mathbf{r})=\delta(\mathbf{r}'-\mathbf{r})+\lambda^{\star}
{\mathsf{g}}(\mathbf{r}',{\mathbf{s}}_{1}){\mathsf{g}}^{\mathsf{H}}(\mathbf{r},{\mathbf{s}}_{1})$ to transform ${\mathsf{Y}}(\mathbf{r})$ as follows:
{\setlength\abovedisplayskip{2pt}
\setlength\belowdisplayskip{2pt}
\begin{equation}\label{MMSE_Filtering_Signal}
\int_{{\mathcal{A}}_{\mathsf{R}}}{\mathscr{K}}_{{\mathsf{Z}}}(\mathbf{r}',\mathbf{r}){\mathsf{Y}}(\mathbf{r}){\rm{d}}\mathbf{r}
=s_2{\mathsf{J}}_2({\mathbf{s}}_{2})\lvert{\mathcal{A}}_{2}\rvert\overline{\mathsf{h}}(\mathbf{r}',{\mathbf{s}}_{2})
+{\mathsf{Z}}_{\mathsf{w}}(\mathbf{r}'),
\end{equation}
}where $\overline{\mathsf{h}}(\mathbf{r}')\triangleq\int_{{\mathcal{A}}_{\mathsf{R}}}{\mathscr{K}}_{{\mathsf{Z}}}(\mathbf{r}',\mathbf{r}){\mathsf{h}}(\mathbf{r},{\mathbf{s}}_{2}){\rm{d}}\mathbf{r}$. Based on Lemma \ref{Lemma_IN_Whiten}, we have ${\mathbbmss{E}}\{{\mathsf{Z}}_{\mathsf{w}}(\mathbf{r}_1){\mathsf{Z}}_{\mathsf{w}}^{\mathsf{H}}(\mathbf{r}_2)\}=
{\sigma}^2\delta(\mathbf{r}_1-\mathbf{r}_2)$. As a result, the model presented in \eqref{MMSE_Filtering_Signal} is similar to that \eqref{Single_User_electric_radiation_field}. Motivated by this, we next design the maximal-ratio combining (MRC)-based detector $\overline{\mathsf{v}}(\mathbf{r}')$ by treating $\overline{\mathsf{h}}(\mathbf{r}')$ as the equivalent channel response, which yields
{\setlength\abovedisplayskip{2pt}
\setlength\belowdisplayskip{2pt}
\begin{equation}
\overline{\mathsf{v}}(\mathbf{r}')=\overline{\mathsf{h}}(\mathbf{r}')
=\int_{{\mathcal{A}}_{\mathsf{R}}}{\mathscr{K}}_{{\mathsf{Z}}}(\mathbf{r}',\mathbf{r}){\mathsf{h}}(\mathbf{r},{\mathbf{s}}_{2}){\rm{d}}\mathbf{r}.
\end{equation}
}The resultant SNR is given by
{\setlength\abovedisplayskip{2pt}
\setlength\belowdisplayskip{2pt}
\begin{equation}\label{Two_User_MMSE_SNR}
\begin{split}
\gamma_2=
\frac{\lvert{\mathsf{J}}_2({\mathbf{s}}_2)\rvert^2\lvert{\mathcal{A}}_2\rvert^2}{\sigma^2}
\int_{{\mathcal{A}}_{\mathsf{R}}}\lvert\overline{\mathsf{h}}(\mathbf{r}')\rvert^2{\rm{d}}{\mathbf{r}}'.
\end{split}
\end{equation}
}A closed-form expression for $\gamma_2$ is given as follows.
\vspace{-5pt}
\begin{theorem}\label{Theorem_MMSE_SNR_First_User}
By first whitening ${\mathsf{Z}}(\mathbf{r})$ with ${\mathscr{K}}_{{\mathsf{Z}}}(\mathbf{r}',\mathbf{r})$ and then employing the MRC detector $\overline{\mathsf{v}}(\mathbf{r}')$, the resultant SNR in decoding $s_2$ can be written as follows:
{\setlength\abovedisplayskip{2pt}
\setlength\belowdisplayskip{2pt}
\begin{equation}\label{Two_User_MMSE_SNR_Closed_Form}
\begin{split}
\gamma_2=\overline{\gamma}_2({\mathsf{a}}_2+\lvert\rho\rvert^2({\lambda^{\star}}^2{\mathsf{a}}_1+2\lambda^{\star})),
\end{split}
\end{equation}
}where $\overline{\gamma}_2=\frac{\lvert{\mathsf{J}}_2({\mathbf{s}}_{2})\rvert^2\lvert{\mathcal{A}}_2\rvert^2k_0^2\eta^2}{4\pi\sigma^2}$ is the transmit SNR of user $2$, ${\mathsf{a}}_{2}=\int_{{\mathcal{A}}_{\mathsf{R}}}\lvert{\mathsf{g}}(\mathbf{r},{\mathbf{s}}_2)\rvert^2{\rm{d}}{\mathbf{r}}$ is the associated channel gain, and $\rho=\int_{{\mathcal{A}}_{\mathsf{R}}}{\mathsf{g}}^{\mathsf{H}}({\mathbf{r}},{\mathbf{s}}_1){\mathsf{g}}({\mathbf{r}},{\mathbf{s}}_2){\rm{d}}{\mathbf{r}}$.  
\end{theorem}
\vspace{-5pt}
\begin{IEEEproof}
Please refer to Appendix \ref{Proof_Theorem_MMSE_SNR_First_User} for more details.
\end{IEEEproof}
After obtaining the capacity-achieving detectors under the SIC order $2\rightarrow1$, we summarize the entire decoding procedure in Table \ref{SIC_Decoding_Two_User} on the bottom of next page for ease of reference. The presented decoding procedure directly extends to the other SIC order $1\rightarrow2$ by exchanging the user indices.
\begin{table}[!t]
\centering
\vspace{5pt}
\setlength{\abovecaptionskip}{0pt}
\begin{tabular}{|c|}
\hline
\begin{minipage}{3in}
\begin{algorithmic}[1]
\State {\footnotesize Whiten the interference-plus-noise term ${\mathsf{Z}}(\mathbf{r})$: 
{\setlength\abovedisplayskip{2pt}
\setlength\belowdisplayskip{2pt}
\begin{equation}
\int_{{\mathcal{A}}_{\mathsf{R}}}{\mathscr{K}}_{{\mathsf{Z}}}(\mathbf{r}',\mathbf{r}){\mathsf{Y}}(\mathbf{r}){\rm{d}}\mathbf{r}
=s_2{\mathsf{J}}_2({\mathbf{s}}_{2})\lvert{\mathcal{A}}_{2}\rvert\overline{\mathsf{h}}(\mathbf{r}',{\mathbf{s}}_{2})
+{\mathsf{Z}}_{\mathsf{w}}(\mathbf{r}')\nonumber
\end{equation}
}}
\State {\footnotesize Use the MRC detector and ML decoder to recover $s_2$:
{\setlength\abovedisplayskip{2pt}
\setlength\belowdisplayskip{2pt}
\begin{equation}
\int_{{\mathcal{A}}_{\mathsf{R}}}{\overline{\mathsf{v}}}^{\mathsf{H}}(\mathbf{r}')\int_{{\mathcal{A}}_{\mathsf{R}}}{\mathscr{K}}_{{\mathsf{Z}}}(\mathbf{r}',\mathbf{r})
{\mathsf{Y}}(\mathbf{r}){\rm{d}}\mathbf{r}{\rm{d}}\mathbf{r}'
\rightarrow s_2 \nonumber
\end{equation}
}}
\State {\footnotesize Employ SIC to subtract $s_2{\mathsf{J}}_2({\mathbf{s}}_{2})\lvert{\mathcal{A}}_{2}\rvert\overline{\mathsf{h}}(\mathbf{r}',{\mathbf{s}}_{2})$ from ${\mathsf{Y}}(\mathbf{r})$}
\State {\footnotesize Use the MRC detector and ML decoder to recover $s_1$:
{\setlength\abovedisplayskip{2pt}
\setlength\belowdisplayskip{2pt}
\begin{equation}
\int_{{\mathcal{A}}_{\mathsf{R}}}{\mathsf{h}}^{\mathsf{H}}(\mathbf{r},{\mathbf{s}}_{1})
(s_1{\mathsf{h}}(\mathbf{r},{\mathbf{s}}_{1}){\mathsf{J}}_1({\mathbf{s}}_{1})\lvert{\mathcal{A}}_{1}\rvert
+{\mathsf{N}}(\mathbf{r})){\rm{d}}\mathbf{r}
\rightarrow s_1 \nonumber
\end{equation}
}}
\end{algorithmic}
\end{minipage}
\\ \hline
\end{tabular}
\caption{SIC decoding for CAPA communications.}
\label{SIC_Decoding_Two_User}
\end{table}
\subsection{Sum-Rate Capacity Analysis}
After obtaining the decoding SNRs for $s_1$ and $s_2$, the achievable rates of user $1$ and user $2$ can be calculated as
{\setlength\abovedisplayskip{2pt}
\setlength\belowdisplayskip{2pt}
\begin{align}
{\mathsf{R}}_1&=\log_2(1+\gamma_1)=\log_2(1+{\overline{\gamma}}_1{\mathsf{a}}_{1}),\label{User1_Rate}\\
{\mathsf{R}}_2&=\log_2(1+\gamma_2)\nonumber\\
&=\log_2(1+\overline{\gamma}_2({\mathsf{a}}_2+\lvert\rho\rvert^2({\lambda^{\star}}^2{\mathsf{a}}_1+2\lambda^{\star}))),\label{User2_Rate}
\end{align}
}respectively. The sum-rate capacity is given as follows.
\vspace{-5pt}
\begin{theorem}\label{Theorem_Sum_Rate_Capacity}
The sum-rate capacity achieved under the SIC order $2\rightarrow1$ is given by
{\setlength\abovedisplayskip{2pt}
\setlength\belowdisplayskip{2pt}
\begin{equation}\label{Sum_Rate_Capacity}
\begin{split}
{\mathsf{C}}=\log_2(1+{\overline{\gamma}}_1{\mathsf{a}}_{1}+{\overline{\gamma}}_2{\mathsf{a}}_{2}+{\overline{\gamma}}_1{\overline{\gamma}}_2{\mathsf{a}}_{1}
{\mathsf{a}}_{2}(1-\lvert\rho_{\mathsf{u}}\rvert^2)),
\end{split}
\end{equation}
}where $\rho_{\mathsf{u}}=\frac{\rho}{\sqrt{{\mathsf{a}}_{1}{\mathsf{a}}_{2}}}$ represents the channel correlation factor between user $1$ and user $2$.  
\end{theorem}
\vspace{-5pt}
\begin{IEEEproof}
Please refer to Appendix \ref{Proof_Theorem_Sum_Rate_Capacity} for more details.
\end{IEEEproof}
Following the same derivation steps in obtaining \eqref{Sum_Rate_Capacity}, we can also obtain the sum-rate capacity achieved under the SIC order $1\rightarrow2$, which is given as follows.
\vspace{-5pt}
\begin{corollary}\label{Corollary_Sum_Rate_Capacity}
Under the SIC order $1\rightarrow2$, the achievable rates of user 1 and user 2 as well as the sum-rate capacity can be expressed as follows:
{\setlength\abovedisplayskip{2pt}
\setlength\belowdisplayskip{2pt}
\begin{align}
{\mathsf{R}}_1&=\log_2(1+\overline{\gamma}_1({\mathsf{a}}_1+\lvert\rho\rvert^2({\lambda^{\bullet}}^2{\mathsf{a}}_2+2\lambda^{\bullet}))),\\
{\mathsf{R}}_2&=\log_2(1+{\overline{\gamma}}_2{\mathsf{a}}_{2}),\\
{\mathsf{C}}&=\log_2(1+{\overline{\gamma}}_1{\mathsf{a}}_{1}+{\overline{\gamma}}_2{\mathsf{a}}_{2}+{\overline{\gamma}}_1{\overline{\gamma}}_2{\mathsf{a}}_{1}
{\mathsf{a}}_{2}(1-\lvert\rho_{\mathsf{u}}\rvert^2)),\label{Sum_Rate_Capacity_1_2}
\end{align}
}where $\lambda^{\bullet}=-\frac{1}{{\mathsf{a}}_{2}}\pm\frac{1}{{\mathsf{a}}_{2}
\sqrt{1+\overline{\gamma}_{2}{\mathsf{a}}_{2}}}$.
\end{corollary}
\vspace{-5pt}
\begin{IEEEproof}
Similar to the proof of Theorem \ref{Theorem_Sum_Rate_Capacity}.
\end{IEEEproof}
By comparing \eqref{Sum_Rate_Capacity} with \eqref{Sum_Rate_Capacity_1_2}, we find the following result.
\vspace{-5pt}
\begin{theorem}
The sum-rate capacity of the considered uplink CAPA-based channel is always the same, i.e., 
{\setlength\abovedisplayskip{2pt}
\setlength\belowdisplayskip{2pt}
\begin{align}
{\mathsf{C}}=\log_2(1+{\overline{\gamma}}_1{\mathsf{a}}_{1}+{\overline{\gamma}}_2{\mathsf{a}}_{2}+{\overline{\gamma}}_1{\overline{\gamma}}_2{\mathsf{a}}_{1}
{\mathsf{a}}_{2}(1-\lvert\rho_{\mathsf{u}}\rvert^2)),\label{Sum_Rate_Capacity_Exact}
\end{align}
}no matter which decoding order is used.
\end{theorem}
\vspace{-5pt}
\vspace{-5pt}
\begin{remark}
By taking a further look at the expression of the sum-rate capacity \eqref{Sum_Rate_Capacity_Exact}, we note that it is determined by the channel gain of each user and the channel correlation factor, all of which are influenced by the receive aperture ${\mathcal{A}}_{\mathsf{R}}$ and the user locations $\{{\mathbf{s}}_k\}_{k=1}^{2}$. This expression applies to an arbitrary aperture regardless of its location, shape, and size.
\end{remark}
\vspace{-5pt}
\subsection{Special Cases}\label{Section: Special Cases}
The sum-rate capacity characterized in \eqref{Sum_Rate_Capacity} or Corollary \ref{Corollary_Sum_Rate_Capacity} applies to an arbitrary aperture regardless of its location, shape, and size, among others. In the sequel, we specialize ${\mathcal{A}}_{\mathsf{R}}$ to several special cases to unveil more system insights.
\subsubsection{Planar CAPAs}
Let us first consider the case where the CAPA is a contiguous-aperture planar array placed on the $x$-$z$ plane with edges parallel to the axes and physical dimensions $L_x$ and $L_z$ along the $x$- and $z$-axes, as shown in \cite[{\figurename} 2(b)]{ouyang2024reactive}. In this case, we have ${\mathcal{A}}_{\mathsf{R}}=\{[x,0,z]|x\in[-\frac{L_x}{2},\frac{L_x}{2}],z\in[-\frac{L_z}{2},\frac{L_z}{2}]\}$ and ${\mathbf{e}}_{{\mathbf{r}}}=[0,1,0]^{\mathsf{T}}$, which yields
{\setlength\abovedisplayskip{2pt}
\setlength\belowdisplayskip{2pt}
\begin{align}
{\mathsf{g}}(\mathbf{r},{\mathbf{s}}_k)=\frac{\sqrt{r_k\Psi_k}{\rm{e}}^{-{\rm{j}}k_0((x-r_k\Phi_k)^2+{\Psi}_k^2+(z-\Theta_k)^2)^{\frac{1}{2}}}}{\sqrt{4\pi} 
((x-r_k\Phi_k)^2+{\Psi}_k^2+(z-\Theta_k)^2)^{\frac{3}{2}}}.
\end{align}
}And the channel gain is calculated as follows.
\vspace{-5pt}
\begin{corollary}
When ${\mathcal{A}}_{\mathsf{R}}=\{[x,0,z]|x\in[-\frac{L_x}{2},\frac{L_x}{2}],z\in[-\frac{L_z}{2},\frac{L_z}{2}]\}$ and ${\mathbf{e}}_{{\mathbf{r}}}=[0,1,0]^{\mathsf{T}}$, the channel gain is give by
{\setlength\abovedisplayskip{2pt}
\setlength\belowdisplayskip{2pt}
\begin{align}\label{CAPA_UPA_Channel_Gain}
{\mathsf{a}}_k=\frac{1}{4\pi}\sum_{x\in{\mathcal{X}}_k}\sum_{z\in{\mathcal{Z}}_k}\arctan\bigg(\frac{xz/\Psi_k}{\sqrt{\Psi_k^2+x^2+z^2}}\bigg)\triangleq 
{\mathsf{a}}_k^{\mathsf{c}},
\end{align}
}where ${\mathcal{X}}_k\triangleq\{\frac{L_x}{2r_k}\pm \Phi_k\}$ and ${\mathcal{Z}}_k\triangleq\{\frac{L_z}{2r_k}\pm \Theta_k\}$.
\end{corollary}
\vspace{-5pt}
\begin{IEEEproof}
Please refer to \cite{liu2023near} for more details.
\end{IEEEproof}
However, deriving a closed-form expression for the correlation factor $\rho_{\mathsf{u}}$ is a challenging task. As is widely known, the correlation factor decreases with the aperture size and $\lim_{\lvert{\mathcal{A}}_{\mathsf{R}}\rvert\rightarrow\infty}\lvert\rho_{\mathsf{u}}\rvert\approx 0$ \cite{liu2024road}. Upon neglecting the channel correlation, we approximate the sum-rate capacity as follows:
{\setlength\abovedisplayskip{2pt}
\setlength\belowdisplayskip{2pt}
\begin{equation}\label{Sum_Rate_Capacity_Appr}
\begin{split}
{\mathsf{C}}&\approx\log_2(1+{\overline{\gamma}}_1{\mathsf{a}}_{1}+{\overline{\gamma}}_2{\mathsf{a}}_{2}+{\overline{\gamma}}_1{\overline{\gamma}}_2{\mathsf{a}}_{1}
{\mathsf{a}}_{2})\\
&=\log_2(1+{\overline{\gamma}}_1{\mathsf{a}}_{1})+\log_2(1+{\overline{\gamma}}_2{\mathsf{a}}_{2}),
\end{split}
\end{equation}
}which serves as an upper bound of the sum-rate capacity. By considering a limiting case where the aperture is infinitely large, i.e., $L_x,L_z\rightarrow\infty$, we have $\lim_{L_x,L_z\rightarrow\infty}{\mathsf{a}}_k^{\mathsf{c}}=\frac{1}{4\pi}\frac{4\pi}{2}=\frac{1}{2}$, and the asymptotic sum-rate capacity is given as follows:
{\setlength\abovedisplayskip{2pt}
\setlength\belowdisplayskip{2pt}
\begin{equation}\label{Sum_Rate_Capacity_CAP}
\begin{split}
{\mathsf{C}}\approx \log_2(1+{\overline{\gamma}}_1/2)+\log_2(1+{\overline{\gamma}}_2/2).
\end{split}
\end{equation}
}
\subsubsection{SPD Planar Arrays}
We then consider a case where the above planar CAPA is partitioned into $M=M_zM_x$ SPD elements, where $M_{x}=2\tilde{M}_x+1$ and $M_{z}=2\tilde{M}_z+1$ denote the number of antenna elements along the $x$- and $z$-axes, as depicted in \cite[{\figurename} 2(a)]{ouyang2024reactive}. The physical dimensions of each element along the $x$- and $z$-axes are indicated by $\sqrt{A}$, and the inter-element distance is denoted as $d$, where $d\geq\sqrt{A}$. In this case, we have $L_x\approx M_xd$, $L_z\approx M_zd$, and ${\mathcal{A}}_{\mathsf{R}}=\{(m_xd+\ell,0,m_zd+\ell)|\ell\in[-\frac{\sqrt{A}}{2},\frac{\sqrt{A}}{2}],m_x\in{\mathcal{M}}_x,m_z\in{\mathcal{M}}_z\}$, where ${\mathcal{M}}_x\triangleq\{0,\pm1,\ldots,\pm\tilde{M}_x\}$ and ${\mathcal{M}}_z\triangleq\{0,\pm1,\ldots,\pm\tilde{M}_z\}$. Under this condition that $\sqrt{A}\ll r_k$ and $d\ll r_k$, we can approximate the corresponding channel gain as follows \cite{ouyang2024reactive}:
{\setlength\abovedisplayskip{2pt}
\setlength\belowdisplayskip{2pt}
\begin{equation}\label{SPD_CAPA_Channel_Gain}
{\mathsf{a}}_k\approx \mu_{\mathsf{oc}}{\mathsf{a}}_k^{\mathsf{c}}\triangleq {\mathsf{a}}_k^{\mathsf{s}},
\end{equation}
}where $\mu_{\mathsf{oc}}\triangleq\frac{A}{d^2}\leq 1$ represents the array occupation ratio. On this basis, we obtain $\lim_{L_x,L_z\rightarrow\infty}{\mathsf{a}}_k^{\mathsf{s}}=\frac{\mu_{\mathsf{oc}}}{2}$, and the asymptotic sum-rate capacity is given as follows:
{\setlength\abovedisplayskip{2pt}
\setlength\belowdisplayskip{2pt}
\begin{equation}\label{Sum_Rate_Capacity_SPD}
\begin{split}
{\mathsf{C}}&\approx \log_2(1+\mu_{\mathsf{oc}}{\overline{\gamma}}_1/2)+\log_2(1+\mu_{\mathsf{oc}}{\overline{\gamma}}_2/2)\\
&\leq \log_2(1+{\overline{\gamma}}_1/2)+\log_2(1+{\overline{\gamma}}_2/2).
\end{split}
\end{equation}
}Comparing \eqref{Sum_Rate_Capacity_CAP} with \eqref{Sum_Rate_Capacity_SPD} yields the following observations.
\vspace{-5pt}
\begin{remark}\label{Remark_CAP_SPD}
The capacity achieved an SPD array converges to that achieved by a CAPA when $\mu_{\mathsf{oc}}=1$. This makes intuitive sense as a CAPA is a special case of an SPD array when the array occupation ratio equals one. 
\end{remark}
\vspace{-5pt}
\section{Numerical Results}
In this section, we conduct computer simulations to analyze the performance of CAPA communications. All simulations utilize planar arrays with all edges parallel to the axes. Unless explicitly stated otherwise, we set the parameters as follows: ${\mathbf{e}}_{{\mathbf{r}}}=[0,1,0]^{\mathsf{T}}$, $\overline{\gamma}_1=30$ dB, $\overline{\gamma}_2=40$ dB, $\theta_1=\theta_2=\frac{\pi}{6}$, $\phi_1=\phi_2=\frac{\pi}{3}$, $r_1=10$ m, $r_2=20$ m, $\lambda = 0.0107$ m, $d=\frac{\lambda}{2}$, $A=\frac{\lambda^2}{4\pi}$, $M_x=M_z$, and $L_x=L_z$.
\begin{figure}[!t]
    \centering
    \subfigbottomskip=0pt
	\subfigcapskip=-5pt
\setlength{\abovecaptionskip}{0pt}
    \subfigure[CAPAs.]
    {
        \includegraphics[height=0.17\textwidth]{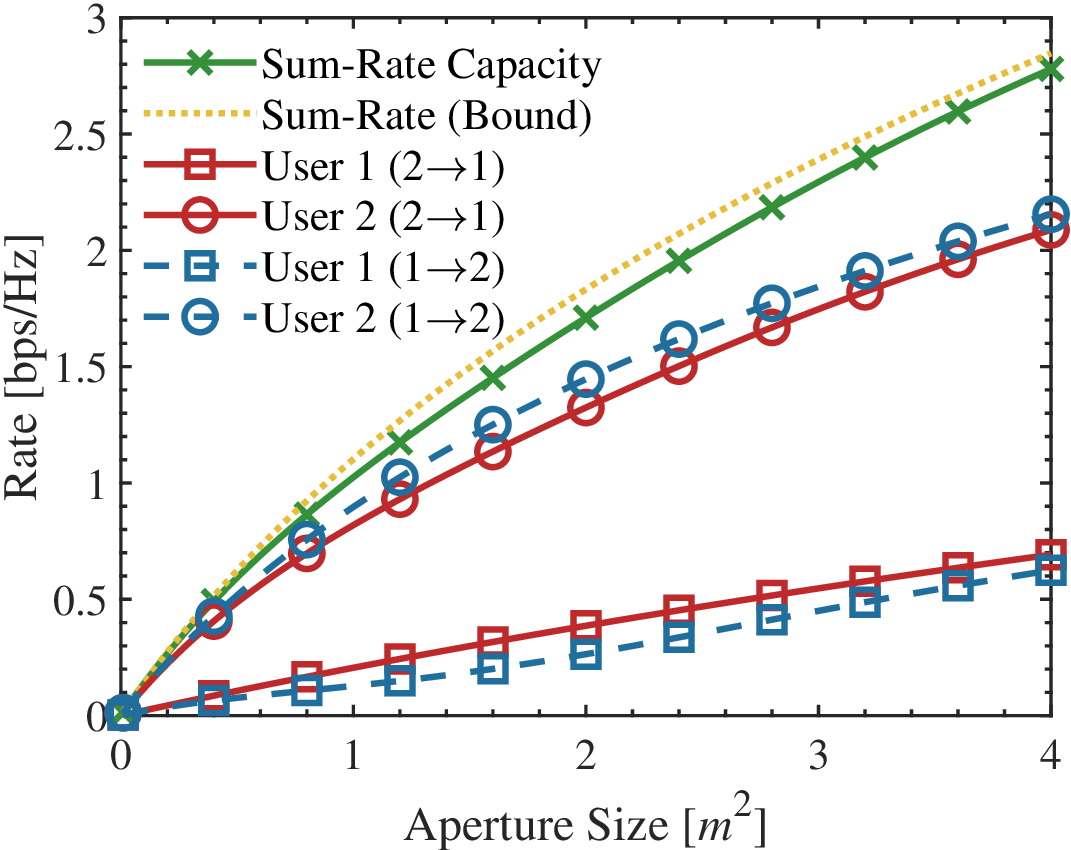}
	   \label{fig2a}	
    }
   \subfigure[SPD arrays.]
    {
        \includegraphics[height=0.17\textwidth]{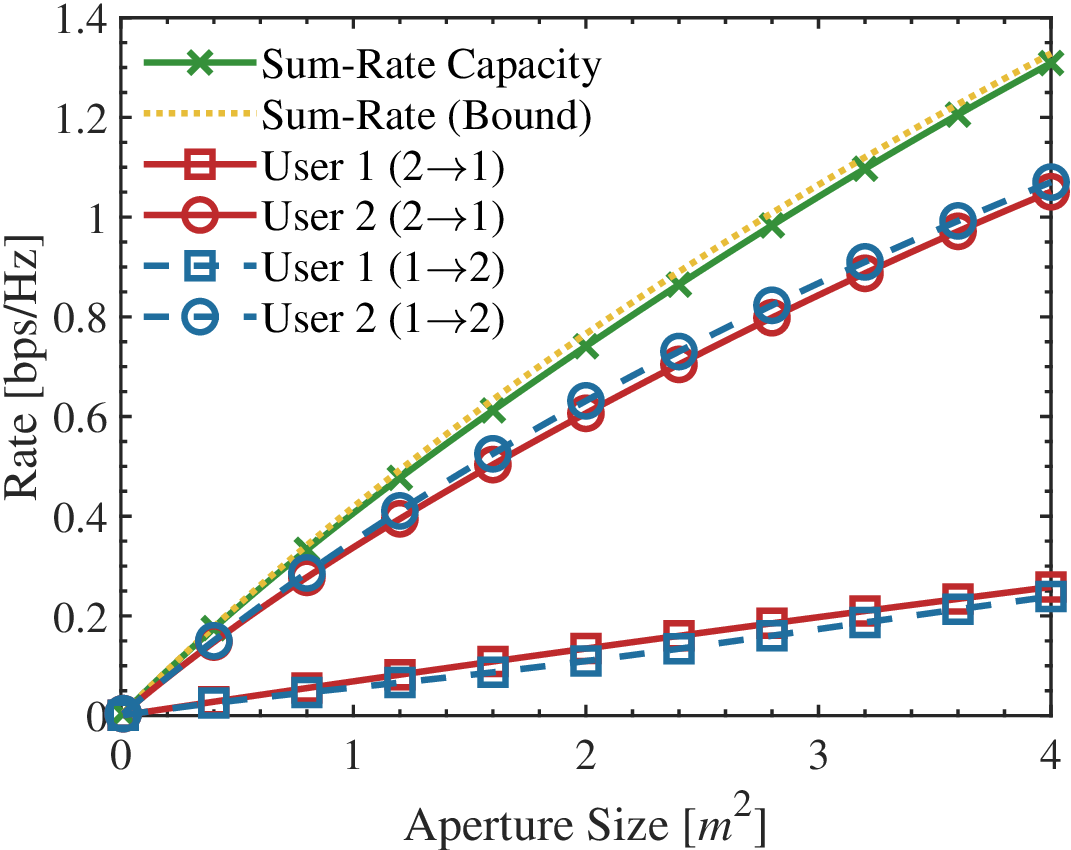}
	   \label{fig2b}	
    }
\caption{Transmission rates achieved by CAPAs and SPD arrays.}
    \label{Figure2}
    \vspace{-10pt}
\end{figure}

{\figurename} {\ref{fig2a}} and {\figurename} {\ref{fig2b}} plot the transmission rates of user $1$ (${\mathsf{R}}_1$) and user $2$ ($\mathsf{R}_2$) under SIC decoding order $1\rightarrow2$ and $2\rightarrow1$, as well as the sum-rate capacity ($\mathsf{C}$) and its upper bound (sum-rate without considering the IUI) achieved by the CAPA and the SPD array, respectively. It can be observed that with the increase in the aperture size $\lvert{\mathcal{A}}_{\mathsf{R}}\rvert$, all presented rates experience an increase. This phenomenon occurs because a larger aperture size leads to a higher channel gain and a lower channel correlation factor, thereby enhancing the dedicated signal while diminishing the IUI. This also causes the sum-rate capacity to approach its upper bound, as derived in \eqref{Sum_Rate_Capacity_CAP} and \eqref{Sum_Rate_Capacity_SPD}, which aligns with the findings illustrated in {\figurename} {\ref{Figure2}}.

\begin{figure}[!t]
    \centering
    \subfigbottomskip=0pt
	\subfigcapskip=-5pt
\setlength{\abovecaptionskip}{0pt}
    \subfigure[Sum-rate capacity vs. $\mu_{\mathsf{oc}}$.]
    {
        \includegraphics[height=0.17\textwidth]{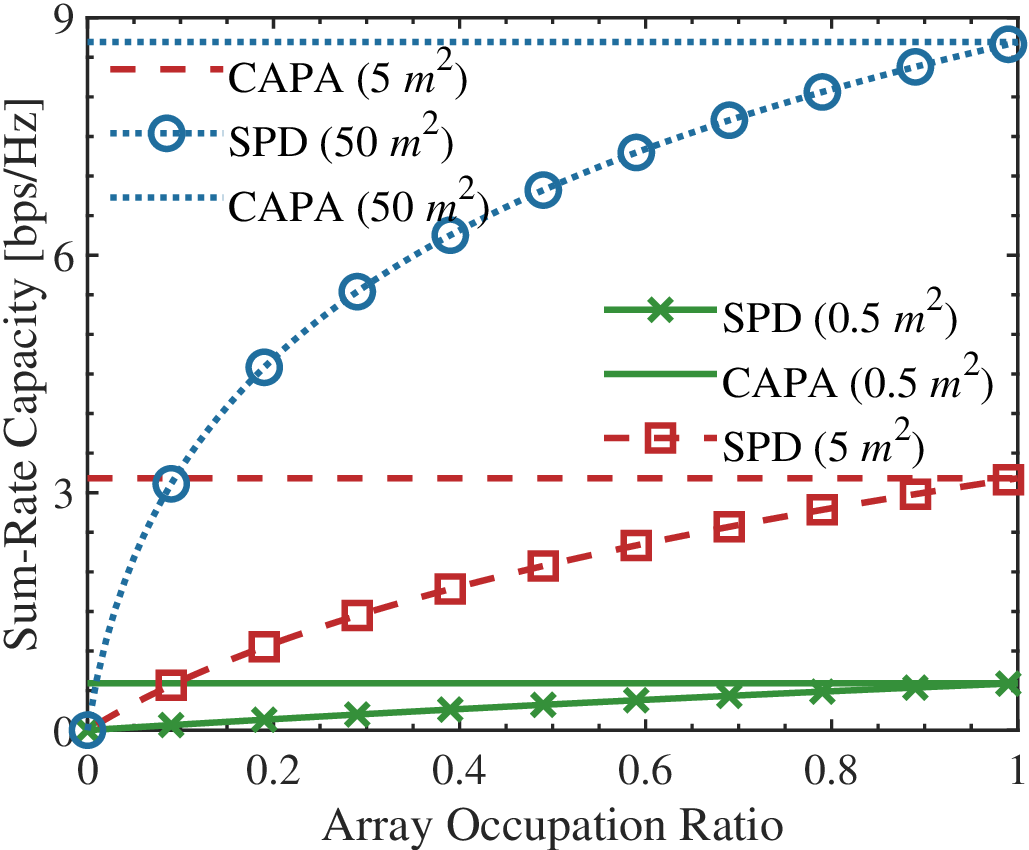}
	   \label{fig3a}	
    }
   \subfigure[Sum-rate capacity vs. $\lvert{\mathcal{A}}_{\mathsf{R}}\rvert$.]
    {
        \includegraphics[height=0.17\textwidth]{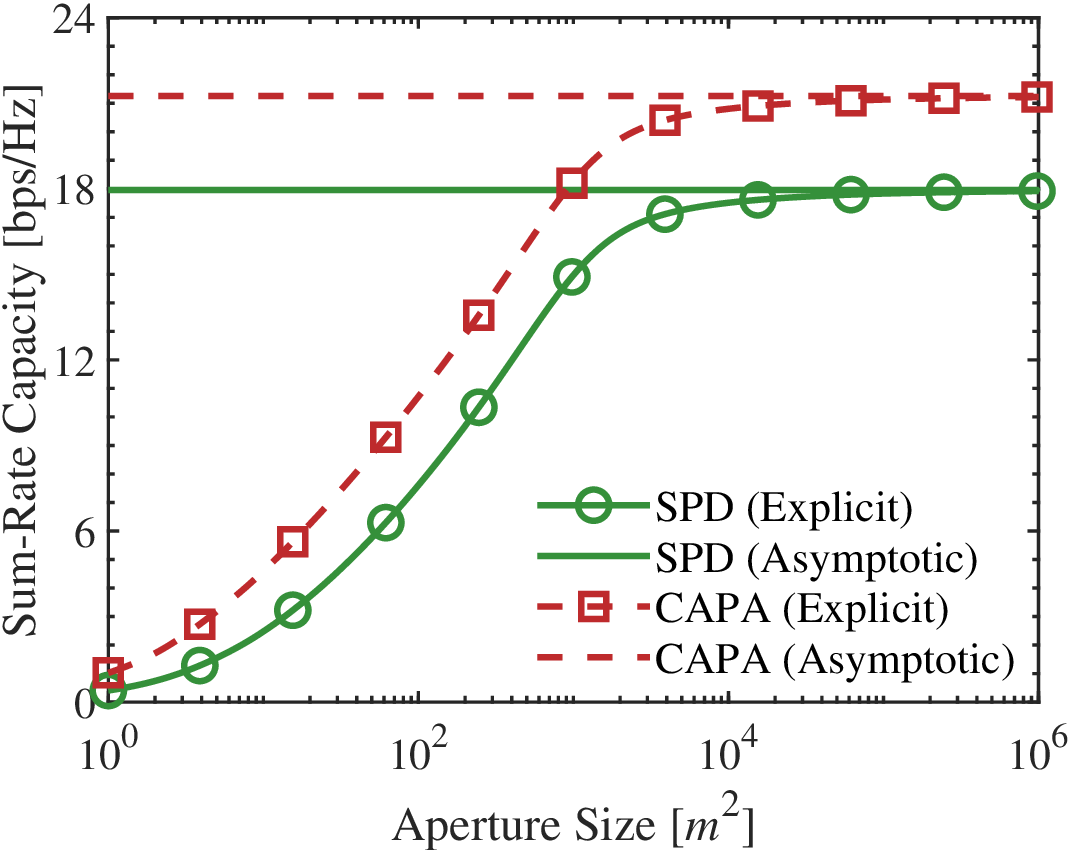}
	   \label{fig3b}	
    }
\caption{Sum-rate capacity of CAPAs and SPD arrays.}
    \label{Figure3}
    \vspace{-10pt}
\end{figure}
Upon comparing {\figurename} {\ref{fig2a}} with {\figurename} {\ref{fig2b}}, we note that the CAPA achieves a higher sum-rate capacity than an SPD array. Our attention then turns to examining the changing trend of the sum-rate capacity achieved by SPD arrays versus the array occupation ratio $\mu_{\mathsf{oc}}$, which is illustrated in {\figurename} {\ref{fig3a}} by fixing $d=\frac{\lambda}{2}$ and changing $A$. We observe that as the array occupation ratio increases, the sum-rate capacity achieved by an SPD array gradually converges to that achieved by a CAPA. This observation further validates the conclusion drawn in Remark \ref{Remark_CAP_SPD}. Additionally, the results depicted in {\figurename} {\ref{fig3a}} suggest that increasing the aperture size can enhance the channel capacity. To delve deeper into the capacity limit as $\lvert{\mathcal{A}}_{\mathsf{R}}\rvert$ approaches infinity, we plot the sum-rate capacity against the aperture size in {\figurename} {\ref{fig3b}}. As expected, the channel capacity converges to a constant value, i.e., the asymptotic value (calculated by \eqref{Sum_Rate_Capacity_CAP} or \eqref{Sum_Rate_Capacity_SPD}), in the larger limit of $\lvert{\mathcal{A}}_{\mathsf{R}}\rvert$, which adheres to the law of energy conservation \cite{liu2023near}.

\begin{figure}[!t]
    \centering
    \subfigbottomskip=0pt
	\subfigcapskip=-5pt
\setlength{\abovecaptionskip}{0pt}
    \subfigure[Capacity region vs. $\mu_{\mathsf{oc}}$. $d=\frac{\lambda}{2}$]
    {
        \includegraphics[height=0.17\textwidth]{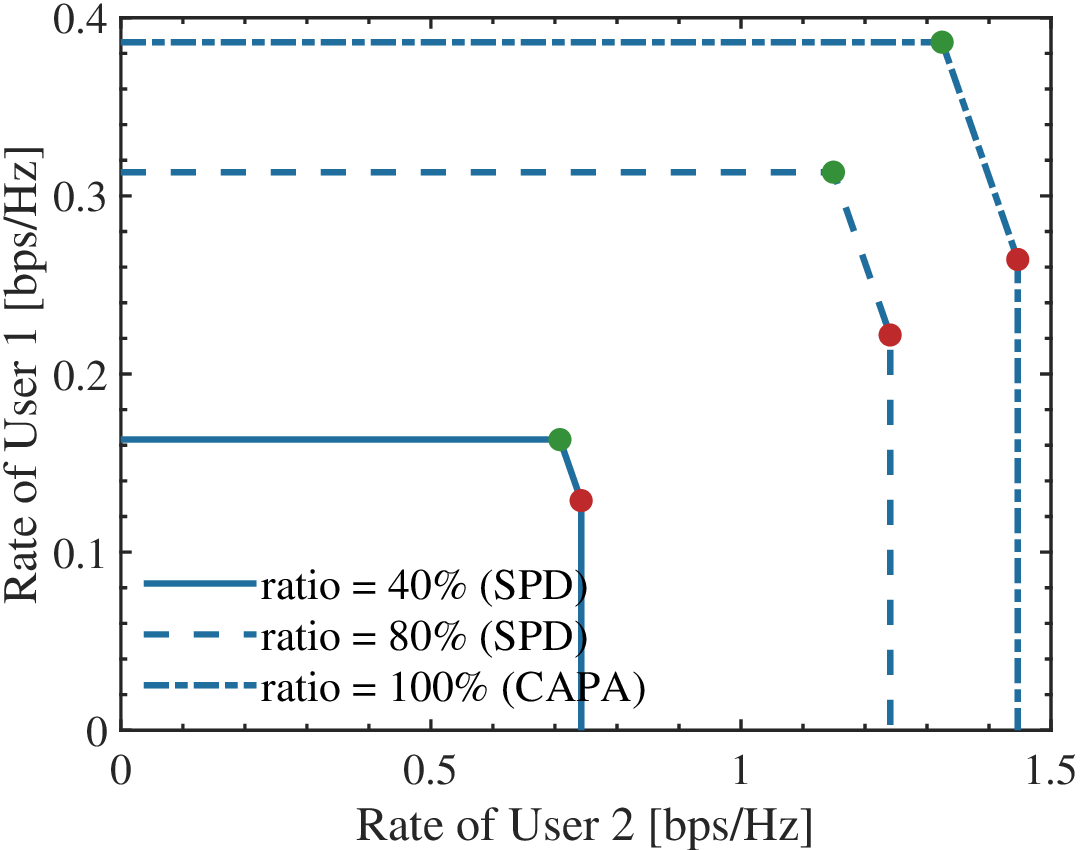}
	   \label{fig4a}	
    }
   \subfigure[Capacity region vs. $\lvert{\mathcal{A}}_{\mathsf{R}}\rvert$.]
    {
        \includegraphics[height=0.17\textwidth]{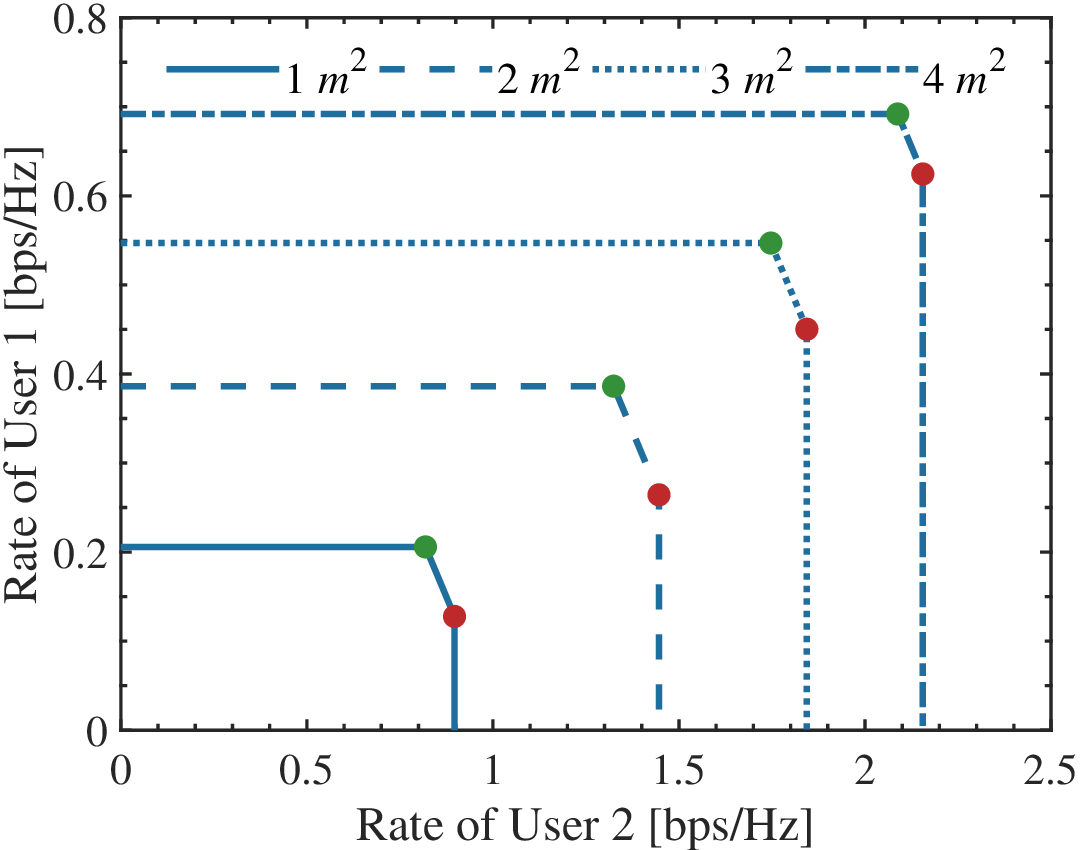}
	   \label{fig4b}	
    }
\caption{Capacity regions of CAPAs and SPD arrays.}
    \label{Figure4}
    \vspace{-10pt}
\end{figure}
{\figurename} {\ref{Figure4}} illustrates the capacity regions achieved by the CAPA and SPD array, where the red and green points represent the achieved rates by \emph{SIC decoding} order $1\rightarrow2$ and $2\rightarrow1$, respectively. The rate tuple on the line segment connecting these two points is achieved by the \emph{time-sharing strategy}. As depicted in {\figurename} {\ref{fig4a}}, the CAPA yields a broader capacity region than the SPD array, and their gap decreases with increasing $\mu_{\mathsf{oc}}$, which is consistent with the results shown in {\figurename} {\ref{fig3a}}. In {\figurename} {\ref{fig4b}}, the capacity regions achieved by the CAPA for various values of $\lvert{\mathcal{A}}_{\mathsf{R}}\rvert$ are illustrated. It is observed that the capacity region gradually extends as $\lvert{\mathcal{A}}_{\mathsf{R}}\rvert$ increases, which transitions from a pentagon to a rectangular shape. This phenomenon occurs because the IUI gradually decreases as the aperture size increases, which aligns with the results shown in {\figurename} {\ref{fig2a}}. These findings underscore the superiority of CAPAs over traditional SPD arrays in terms of channel capacity.
\section{Conclusion} 
We have proposed an analytically tractable framework for CAPA-based wireless communications. Utilizing this framework, we derived the channel capacity for an uplink CAPA-based system, along with the associated decoding principle. We have demonstrated through both theoretical analyses and numerical simulations that CAPA exhibits a higher sum-rate capacity and a broader capacity region compared to the traditional SPD array. These results highlight CAPA’s potential as a promising transmission paradigm for future wireless networks.
\begin{appendix}
\subsection{Proof of Lemma \ref{Lemma_Noise_Distribution}}\label{Proof_Lemma_Noise_Distribution}
The mean of $\int_{{\mathcal{A}}_{\mathsf{R}}}{\mathsf{v}}^{\mathsf{H}}(\mathbf{r}){\mathsf{N}}(\mathbf{r}){\rm{d}}{\mathbf{r}}$ is calculated as follows:
{\setlength\abovedisplayskip{2pt}
\setlength\belowdisplayskip{2pt}
\begin{equation}
\begin{split}
{\mathbbmss{E}}\left\{\int_{{\mathcal{A}}_{\mathsf{R}}}{\mathsf{v}}^{\mathsf{H}}(\mathbf{r}){\mathsf{N}}(\mathbf{r}){\rm{d}}{\mathbf{r}}\right\}
=\int_{{\mathcal{A}}_{\mathsf{R}}}{\mathsf{v}}^{\mathsf{H}}(\mathbf{r}){\mathbbmss{E}}\{{\mathsf{N}}(\mathbf{r})\}{\rm{d}}{\mathbf{r}},
\end{split}
\end{equation}
}which, together with the fact that ${\mathbbmss{E}}\{{\mathsf{N}}(\mathbf{r})\}=0$, yields ${\mathbbmss{E}}\{\int_{{\mathcal{A}}_{\mathsf{R}}}{\mathsf{v}}^{\mathsf{H}}(\mathbf{r}){\mathsf{N}}(\mathbf{r}){\rm{d}}{\mathbf{r}}\}=0$. The variance is given by
{\setlength\abovedisplayskip{2pt}
\setlength\belowdisplayskip{2pt}
\begin{equation}\label{Noise_Field_Covariance}
\begin{split}
&{\mathbbmss{E}}\left\{\int_{{\mathcal{A}}_{\mathsf{R}}}{\mathsf{v}}^{\mathsf{H}}(\mathbf{r}){\mathsf{N}}(\mathbf{r}){\rm{d}}{\mathbf{r}}
\int_{{\mathcal{A}}_{\mathsf{R}}}{\mathsf{v}}(\mathbf{r}'){\mathsf{N}}^{\mathsf{H}}(\mathbf{r}'){\rm{d}}{\mathbf{r}}'\right\}\\
&=\int_{{\mathcal{A}}_{\mathsf{R}}}\int_{{\mathcal{A}}_{\mathsf{R}}}{\mathsf{v}}^{\mathsf{H}}(\mathbf{r})
{\mathsf{v}}(\mathbf{r}')
{\mathbbmss{E}}\{{\mathsf{N}}(\mathbf{r}){\mathsf{N}}^{\mathsf{H}}(\mathbf{r}')\}{\rm{d}}{\mathbf{r}}{\rm{d}}{\mathbf{r}}'\\
&=\int_{{\mathcal{A}}_{\mathsf{R}}}{\mathsf{v}}(\mathbf{r}')\int_{{\mathcal{A}}_{\mathsf{R}}}{\mathsf{v}}^{\mathsf{H}}(\mathbf{r})
\sigma^2\delta(\mathbf{r}-\mathbf{r}'){\rm{d}}{\mathbf{r}}{\rm{d}}{\mathbf{r}}'.
\end{split}
\end{equation}
}Using the fact that $\int_{{\mathcal{A}}_{\mathsf{R}}}\delta({\mathbf{x}}-{\mathbf{x}}_0)f(\mathbf{x}){\rm{d}}{\mathbf{x}}=f({\mathbf{x}}_0)$, where $f(\cdot)$ is an arbitrary function defined on ${\mathcal{A}}_{\mathsf{R}}$, we obtain
{\setlength\abovedisplayskip{2pt}
\setlength\belowdisplayskip{2pt}
\begin{equation}
\eqref{Noise_Field_Covariance}=\int_{{\mathcal{A}}_{\mathsf{R}}}{\mathsf{v}}(\mathbf{r}')(\sigma^2{\mathsf{v}}^{\mathsf{H}}(\mathbf{r}')){\rm{d}}{\mathbf{r}}'=\sigma^2\int_{{\mathcal{A}}_{\mathsf{R}}}\lvert{\mathsf{v}}(\mathbf{r})\rvert^2{\rm{d}}{\mathbf{r}}.
\end{equation}
}The lemma is thus proved.
\subsection{Proof of Lemma \ref{Lemma_IN_Invertible}}\label{Proof_Lemma_IN_Invertible}
By virtue of changing the order of integration in $\int_{{\mathcal{A}}_{\mathsf{R}}}\overline{\mathscr{K}}_{{\mathsf{Z}}}(\mathbf{r}'',\mathbf{r}')
\int_{{\mathcal{A}}_{\mathsf{R}}}{\mathscr{K}}_{{\mathsf{Z}}}(\mathbf{r}',\mathbf{r}){\mathsf{u}}(\mathbf{r}){\rm{d}}\mathbf{r}{\rm{d}}\mathbf{r}'$, we obtain
{\setlength\abovedisplayskip{2pt}
\setlength\belowdisplayskip{2pt}
\begin{equation}\label{Proof_Lemma_IN_Invertible_Step1}
\begin{split}
&\int_{{\mathcal{A}}_{\mathsf{R}}}\overline{\mathscr{K}}_{{\mathsf{Z}}}(\mathbf{r}'',\mathbf{r}')
\int_{{\mathcal{A}}_{\mathsf{R}}}{\mathscr{K}}_{{\mathsf{Z}}}(\mathbf{r}',\mathbf{r}){\mathsf{u}}(\mathbf{r})
{\rm{d}}\mathbf{r}{\rm{d}}\mathbf{r}'\\
&=\int_{{\mathcal{A}}_{\mathsf{R}}}{\mathsf{u}}(\mathbf{r})\int_{{\mathcal{A}}_{\mathsf{R}}}
\overline{\mathscr{K}}_{{\mathsf{Z}}}(\mathbf{r}'',\mathbf{r}'){\mathscr{K}}_{{\mathsf{Z}}}(\mathbf{r}',\mathbf{r})
{\rm{d}}\mathbf{r}'{\rm{d}}\mathbf{r}.
\end{split}
\end{equation}
}By inserting ${\mathscr{K}}_{{\mathsf{Z}}}(\mathbf{r}',\mathbf{r})=\delta(\mathbf{r}'-\mathbf{r})+\lambda
{\mathsf{g}}(\mathbf{r}',{\mathbf{s}}_{1}){\mathsf{g}}^{\mathsf{H}}(\mathbf{r},{\mathbf{s}}_{1})$ and $\overline{\mathscr{K}}_{{\mathsf{Z}}}(\mathbf{r}',\mathbf{r})=\delta(\mathbf{r}'-\mathbf{r})-\overline{\lambda}
{\mathsf{g}}(\mathbf{r}',{\mathbf{s}}_{1}){\mathsf{g}}^{\mathsf{H}}(\mathbf{r},{\mathbf{s}}_{1})$ into \eqref{Proof_Lemma_IN_Invertible_Step1} as well as calculating the resultant integral using the fact that $\int_{{\mathcal{A}}_{\mathsf{R}}}\delta({\mathbf{x}}-{\mathbf{x}}_0)f(\mathbf{x}){\rm{d}}{\mathbf{x}}=f({\mathbf{x}}_0)$ and ${\mathsf{a}}_{1}=\int_{{\mathcal{A}}_{\mathsf{R}}}\lvert{\mathsf{g}}(\mathbf{r},{\mathbf{s}}_1)\rvert^2{\rm{d}}{\mathbf{r}}$, we obtain 
{\setlength\abovedisplayskip{2pt}
\setlength\belowdisplayskip{2pt}
\begin{equation}\label{Proof_Lemma_IN_Invertible_Step2}
\begin{split}
&\int_{{\mathcal{A}}_{\mathsf{R}}}
\overline{\mathscr{K}}_{{\mathsf{Z}}}(\mathbf{r}'',\mathbf{r}'){\mathscr{K}}_{{\mathsf{Z}}}(\mathbf{r}',\mathbf{r})
{\rm{d}}\mathbf{r}'=\delta({\mathbf{r}}''-{\mathbf{r}})\\
&-(\overline{\lambda}-\lambda+\lambda\overline{\lambda}{\mathsf{a}}_1){\mathsf{g}}(\mathbf{r}'',\mathbf{s}_1){\mathsf{g}}^{\mathsf{H}}(\mathbf{r}',\mathbf{s}_1).
\end{split}
\end{equation}
}Inserting $\overline{\lambda}=\frac{\lambda}{1+\lambda {\mathsf{a}}_{1}}$ into $\overline{\lambda}-\lambda+\lambda\overline{\lambda}{\mathsf{a}}_1$ gives
{\setlength\abovedisplayskip{2pt}
\setlength\belowdisplayskip{2pt}
\begin{equation}
\overline{\lambda}-\lambda+\lambda\overline{\lambda}{\mathsf{a}}_1=\overline{\lambda}(1+\lambda{\mathsf{a}}_1)-\lambda=0, 
\end{equation}
}which, together with the fact in \eqref{Proof_Lemma_IN_Invertible_Step2}, yields
{\setlength\abovedisplayskip{2pt}
\setlength\belowdisplayskip{2pt}
\begin{equation}\label{Proof_Lemma_IN_Invertible_Step3}
\begin{split}
\int_{{\mathcal{A}}_{\mathsf{R}}}
\overline{\mathscr{K}}_{{\mathsf{Z}}}(\mathbf{r}'',\mathbf{r}'){\mathscr{K}}_{{\mathsf{Z}}}(\mathbf{r}',\mathbf{r})
{\rm{d}}\mathbf{r}'=\delta({\mathbf{r}}''-{\mathbf{r}}).
\end{split}
\end{equation}
}Substituting \eqref{Proof_Lemma_IN_Invertible_Step3} into \eqref{Proof_Lemma_IN_Invertible_Step1} gives
{\setlength\abovedisplayskip{2pt}
\setlength\belowdisplayskip{2pt}
\begin{equation}\label{Proof_Lemma_IN_Invertible_Step4}
\begin{split}
&\int_{{\mathcal{A}}_{\mathsf{R}}}\overline{\mathscr{K}}_{{\mathsf{Z}}}(\mathbf{r}'',\mathbf{r}')
\int_{{\mathcal{A}}_{\mathsf{R}}}{\mathscr{K}}_{{\mathsf{Z}}}(\mathbf{r}',\mathbf{r}){\mathsf{u}}(\mathbf{r})
{\rm{d}}\mathbf{r}{\rm{d}}\mathbf{r}'\\
&=\int_{{\mathcal{A}}_{\mathsf{R}}}{\mathsf{u}}(\mathbf{r})\delta({\mathbf{r}}''-{\mathbf{r}}){\rm{d}}\mathbf{r}={\mathsf{u}}(\mathbf{r}'').
\end{split}
\end{equation}
}The lemma is thus proved.
\subsection{Proof of Lemma \ref{Lemma_IN_Whiten}}\label{Proof_Lemma_IN_Whiten}
The autocorrelation function of ${\mathsf{Z}}_{\mathsf{w}}(\cdot)$ satisfies
{\setlength\abovedisplayskip{2pt}
\setlength\belowdisplayskip{2pt}
\begin{align}
&{\mathbbmss{E}}\{{\mathsf{Z}}_{\mathsf{w}}(\mathbf{r}_1){\mathsf{Z}}_{\mathsf{w}}^{\mathsf{H}}(\mathbf{r}_2)\}\nonumber\\
&=\int_{{\mathcal{A}}_{\mathsf{R}}}\int_{{\mathcal{A}}_{\mathsf{R}}}{\mathscr{K}}_{{\mathsf{Z}}}(\mathbf{r}_1,\mathbf{r}){\mathscr{K}}_{{\mathsf{Z}}}^{\mathsf{H}}(\mathbf{r}_2,\mathbf{r}')
{\mathbbmss{E}}\{{\mathsf{Z}}(\mathbf{r}){\mathsf{Z}}^{\mathsf{H}}(\mathbf{r}')\}{\rm{d}}\mathbf{r}{\rm{d}}\mathbf{r}'\nonumber\\
&=\sigma^2\int_{{\mathcal{A}}_{\mathsf{R}}}{\mathscr{K}}_{{\mathsf{Z}}}^{\mathsf{H}}(\mathbf{r}_2,\mathbf{r}')
\int_{{\mathcal{A}}_{\mathsf{R}}}{\mathscr{K}}_{{\mathsf{Z}}}(\mathbf{r}_1,\mathbf{r})
{\mathscr{R}}_{{\mathsf{Z}}{\mathsf{Z}}}(\mathbf{r},\mathbf{r}'){\rm{d}}\mathbf{r}{\rm{d}}\mathbf{r}'.
\label{Proof_Lemma_IN_Whiten_Step1}
\end{align}
}By substituting ${\mathscr{K}}_{{\mathsf{Z}}}(\mathbf{r}',\mathbf{r})=\delta(\mathbf{r}'-\mathbf{r})+\lambda
{\mathsf{g}}(\mathbf{r}',{\mathbf{s}}_{1}){\mathsf{g}}^{\mathsf{H}}(\mathbf{r},{\mathbf{s}}_{1})$ and \eqref{Autocorrelatioon_Interference_Noise} into \eqref{Proof_Lemma_IN_Whiten_Step1} and then calculating the resultant double integral with respect to $\mathbf{r}$, we obtain
{\setlength\abovedisplayskip{2pt}
\setlength\belowdisplayskip{2pt}
\begin{equation}\label{Denominator_Calculation_Forward_1}
\begin{split}
&\int_{{\mathcal{A}}_{\mathsf{R}}}{\mathscr{K}}_{{\mathsf{Z}}}(\mathbf{r}_1,\mathbf{r})
{\mathscr{R}}_{{\mathsf{Z}}{\mathsf{Z}}}(\mathbf{r},\mathbf{r}'){\rm{d}}\mathbf{r}=\sigma^2(\delta({\mathbf{r}}_1-{\mathbf{r}}')\\
&+(\overline{\gamma}_1+\lambda+\lambda\overline{\gamma}_1{\mathsf{a}}_1){\mathsf{g}}(\mathbf{r}_1,\mathbf{s}_1){\mathsf{g}}^{\mathsf{H}}(\mathbf{r}',\mathbf{s}_1)).
\end{split}
\end{equation}
}We next calculate the integral in terms of ${{\mathbf{r}}}'$, which yields
{\setlength\abovedisplayskip{2pt}
\setlength\belowdisplayskip{2pt}
\begin{equation}\label{Denominator_Calculation_Forward_2}
\begin{split}
&\int_{\mathcal{A}_{\mathsf{S}}}\eqref{Denominator_Calculation_Forward_1}
\times(\delta(\mathbf{r}_2-\mathbf{r}')+\lambda
{\mathsf{g}}^{\mathsf{H}}(\mathbf{r}_2,{\mathbf{s}}_{1}){\mathsf{g}}(\mathbf{r}',{\mathbf{s}}_{1})){\rm{d}}{{\mathbf{r}}'}\\
&=\sigma^2(\delta({\mathbf{r}}_1-{\mathbf{r}}_2)+{\mathsf{g}}(\mathbf{r}_1,\mathbf{s}_1){\mathsf{g}}^{\mathsf{H}}(\mathbf{r}_2,\mathbf{s}_1)\times(2\lambda+\overline{\gamma}_1\\
&+2\lambda\overline{\gamma}_1{{\mathsf{a}}_{1}}+\lambda^2{{\mathsf{a}}_{1}}
+\overline{\gamma}_1\lambda^2{{\mathsf{a}}_{1}^2})).
\end{split}
\end{equation}
}It is readily shown that the solutions to the equation $2\lambda+\overline{\gamma}_1+2\lambda\overline{\gamma}_1{{\mathsf{a}}_{1}}+\lambda^2{{\mathsf{a}}_{1}}
+\overline{\gamma}_1\lambda^2{{\mathsf{a}}_{1}^2}=0$ are given by $\lambda=-\frac{1}{{\mathsf{a}}_{1}}\pm\frac{1}{{\mathsf{a}}_{1}
\sqrt{1+\overline{\gamma}_{1}{\mathsf{a}}_{1}}}$. Hence, setting $\lambda=-\frac{1}{{\mathsf{a}}_{1}}\pm\frac{1}{{\mathsf{a}}_{1}
\sqrt{1+\overline{\gamma}_{1}{\mathsf{a}}_{1}}}$ gives ${\mathbbmss{E}}\{{\mathsf{Z}}_{\mathsf{w}}(\mathbf{r}_1){\mathsf{Z}}_{\mathsf{w}}^{\mathsf{H}}(\mathbf{r}_2)\}=\sigma^2\delta({\mathbf{r}}_1-{\mathbf{r}}_2)$. The lemma is thus proved.
\subsection{Proof of Theorem \ref{Theorem_MMSE_SNR_First_User}}\label{Proof_Theorem_MMSE_SNR_First_User}
By definition, $\overline{\mathsf{h}}(\mathbf{r}')$ can be calculated as follows:
{\setlength\abovedisplayskip{2pt}
\setlength\belowdisplayskip{2pt}
\begin{align}
\overline{\mathsf{h}}(\mathbf{r}')&=
\int_{{\mathcal{A}}_{\mathsf{R}}}(\delta(\mathbf{r}'-\mathbf{r})+\lambda^{\star}
{\mathsf{g}}(\mathbf{r}',{\mathbf{s}}_{1}){\mathsf{g}}^{\mathsf{H}}(\mathbf{r},{\mathbf{s}}_{1})){\mathsf{h}}(\mathbf{r},{\mathbf{s}}_{2}){\rm{d}}\mathbf{r}\nonumber\\
&={{\rm{j}}k_0\eta}/{\sqrt{4\pi}}({\mathsf{g}}(\mathbf{r}',{\mathbf{s}}_{2})+\lambda^{\star}{\mathsf{g}}(\mathbf{r}',{\mathbf{s}}_{1})\rho).
\label{Proof_Theorem_MMSE_SNR_First_User_Step1}
\end{align}
}Inserting \eqref{Proof_Theorem_MMSE_SNR_First_User_Step1} into \eqref{Two_User_MMSE_SNR} gives
{\setlength\abovedisplayskip{2pt}
\setlength\belowdisplayskip{2pt}
\begin{equation}\label{Proof_Theorem_MMSE_SNR_First_User_Step2}
\begin{split}
\gamma_2&=\overline{\gamma}_2
\int_{{\mathcal{A}}_{\mathsf{R}}}\lvert{\mathsf{g}}(\mathbf{r}',{\mathbf{s}}_{2})+\lambda^{\star}\rho{\mathsf{g}}(\mathbf{r}',{\mathbf{s}}_{1})\rvert^2{\rm{d}}{\mathbf{r}}'\\
&=\overline{\gamma}_2\int_{{\mathcal{A}}_{\mathsf{R}}}\lvert{\mathsf{g}}(\mathbf{r}',{\mathbf{s}}_{2})\rvert^2{\rm{d}}{\mathbf{r}}'+\overline{\gamma}_2
\int_{{\mathcal{A}}_{\mathsf{R}}}\lvert{\mathsf{g}}(\mathbf{r}',{\mathbf{s}}_{2})\rvert^2{\rm{d}}{\mathbf{r}}'\\
&\times {\lambda^{\star}}^2\lvert\rho\rvert^2+2\overline{\gamma}_2\lambda^{\star}\Re\left\{\rho\int_{{\mathcal{A}}_{\mathsf{R}}}{\mathsf{g}}({\mathbf{r}},{\mathbf{s}}_1){\mathsf{g}}^{\mathsf{H}}({\mathbf{r}},{\mathbf{s}}_2){\rm{d}}{\mathbf{r}}\right\}\\
&=\overline{\gamma}_2({\mathsf{a}}_2+\lvert\rho\rvert^2({\lambda^{\star}}^2{\mathsf{a}}_1+2\lambda^{\star})).
\end{split}
\end{equation}
}The theorem is thus proved.
\subsection{Proof of Theorem \ref{Theorem_Sum_Rate_Capacity}}\label{Proof_Theorem_Sum_Rate_Capacity}
Since $\lambda=\lambda^{\star}$ is the solution to $2\lambda+\overline{\gamma}_1+2\lambda\overline{\gamma}_1{{\mathsf{a}}_{1}}+\lambda^2{{\mathsf{a}}_{1}}
+\overline{\gamma}_1\lambda^2{{\mathsf{a}}_{1}^2}=0$, we have
{\setlength\abovedisplayskip{2pt}
\setlength\belowdisplayskip{2pt}
\begin{equation}
\begin{split}
2\lambda^{\star}+{\lambda^{\star}}^2{{\mathsf{a}}_{1}}&=-\overline{\gamma}_1-2\lambda^{\star}\overline{\gamma}_1{{\mathsf{a}}_{1}}
-\overline{\gamma}_1{\lambda^{\star}}^2{{\mathsf{a}}_{1}^2}\\
&=-\overline{\gamma}_1(1+\lambda^{\star}{\mathsf{a}}_{1})^2,
\end{split}
\end{equation}
}which, together with \eqref{User2_Rate}, yields
{\setlength\abovedisplayskip{2pt}
\setlength\belowdisplayskip{2pt}
\begin{align}\label{Proof_Theorem_Sum_Rate_Capacity_Step1}
{\mathsf{R}}_2
=\log_2(1+\overline{\gamma}_2({\mathsf{a}}_2-\lvert\rho\rvert^2\overline{\gamma}_1(1+\lambda^{\star}{\mathsf{a}}_{1})^2)).
\end{align}
}Furthermore, inserting $\lambda^{\star}=-\frac{1}{{\mathsf{a}}_{1}}\pm\frac{1}{{\mathsf{a}}_{1}
\sqrt{1+\overline{\gamma}_{1}{\mathsf{a}}_{1}}}$ into \eqref{Proof_Theorem_Sum_Rate_Capacity_Step1} gives
{\setlength\abovedisplayskip{2pt}
\setlength\belowdisplayskip{2pt}
\begin{align}\label{Proof_Theorem_Sum_Rate_Capacity_Step2}
{\mathsf{R}}_2
=\log_2(1+\overline{\gamma}_2({\mathsf{a}}_2-\lvert\rho\rvert^2\overline{\gamma}_1/(1+\overline{\gamma}_{1}{\mathsf{a}}_{1}))).
\end{align}
}Substituting \eqref{User1_Rate} and \eqref{Proof_Theorem_Sum_Rate_Capacity_Step2} into ${\mathsf{C}}={\mathsf{R}}_1+{\mathsf{R}}_2$ gives
{\setlength\abovedisplayskip{2pt}
\setlength\belowdisplayskip{2pt}
\begin{equation}
\begin{split}
{\mathsf{C}}&=\log_2(1+\overline{\gamma}_{1}{\mathsf{a}}_{1}+\overline{\gamma}_2({\mathsf{a}}_2(1+\overline{\gamma}_{1}{\mathsf{a}}_{1})-\lvert\rho\rvert^2\overline{\gamma}_1))\\
&=\log_2(1+{\overline{\gamma}}_1{\mathsf{a}}_{1}+{\overline{\gamma}}_2{\mathsf{a}}_{2}+{\overline{\gamma}}_1{\overline{\gamma}}_2{\mathsf{a}}_{1}
{\mathsf{a}}_{2}(1-\lvert\rho_{\mathsf{u}}\rvert^2)).
\end{split}
\end{equation}
}The theorem is proved.
\end{appendix}
\bibliographystyle{IEEEtran}
\bibliography{mybib}
\end{document}